\documentclass[aps,prd,showpacs,nofootinbib,amsmath,superscriptaddress]{revtex4}

\usepackage{bm}
\usepackage{graphicx}

\newcommand{\de}{{\rm d}}

\begin{document}

\title{Azimuthal asymmetries for hadron distributions inside a jet in hadronic collisions}

\author{Umberto~D'Alesio}
 \email{umberto.dalesio@ca.infn.it}
 \affiliation{Dipartimento di Fisica, Universit\`a di Cagliari, Cittadella Universitaria,
 I-09042 Monserrato (CA), Italy}
 \affiliation{Istituto Nazionale di Fisica Nucleare, Sezione di Cagliari, C.P. 170,
 I-09042 Monserrato (CA), Italy}

\author{Francesco~Murgia}
 \email{francesco.murgia@ca.infn.it}
 \affiliation{Istituto Nazionale di Fisica Nucleare, Sezione di Cagliari, C.P. 170,
 I-09042 Monserrato (CA), Italy}

\author{Cristian~Pisano}
 \email{cristian.pisano@ca.infn.it}
 \affiliation{Dipartimento di Fisica, Universit\`a di Cagliari, Cittadella Universitaria,
 I-09042 Monserrato (CA), Italy}
 \affiliation{Istituto Nazionale di Fisica Nucleare, Sezione di Cagliari, C.P. 170,
 I-09042 Monserrato (CA), Italy}

\date{\today}

\begin{abstract}
Using a generalized parton model approach including spin and intrinsic
parton motion effects, and assuming the validity of factorization for
large-$p_T$ jet production in hadronic collisions, we study the azimuthal
distribution around the jet axis of leading unpolarized or (pseudo)scalar hadrons,
namely pions, produced in the jet fragmentation process. We identify the observable
leading-twist azimuthal asymmetries for the unpolarized
and single-polarized case related to quark and gluon-originated jets.
We account for all physically allowed combinations of the transverse momentum  dependent
(TMD) parton distribution and fragmentation functions, with
special attention to the Sivers, Boer-Mulders, and transversity
quark distributions, and to the Collins fragmentation function for quarks
(and to the analogous functions for gluons).
For each of these effects we evaluate,
at central and forward rapidities and for
kinematical configurations accessible at BNL-RHIC,
the corresponding potentially maximized asymmetry
(for $\pi^+$ production), obtained by saturating
natural positivity bounds (and the Soffer bound for transversity)
for the distribution and fragmentation functions involved
and summing additively all partonic contributions.
We then estimate, for both neutral and charged pions,
the asymmetries involving TMD functions for
which parameterizations are available.
We also study the role of the different mechanisms, and the
corresponding transverse single spin asymmetries,
for large-$p_T$ inclusive jet production.
\end{abstract}


\maketitle

\section{\label{intro} Introduction}
Transverse single-spin and azimuthal asymmetries in high-energy hadronic
reactions have raised a lot of interest in the last years
(see e.g.~Refs.~\cite{D'Alesio:2007jt,Barone:2010ef} and references therein).
Huge spin asymmetries have been measured in the inclusive forward
production of pions in high-energy $pp$ collisions at moderately large transverse momentum.
The general trend of the early pioneer measurements of the E704 Collaboration at
Fermilab~\cite{Adams:1991cs,Adams:1991ru}
has been recently confirmed at much larger center of mass (c.m.) energies at
the Relativistic Heavy Ion Collider (RHIC) at Brookhaven National Laboratories (BNL)
in similar kinematical configurations~\cite{Adams:2003fx,Abelev:2008qb}.
A surprisingly large transverse polarization of $\Lambda$ hyperons produced
in the forward region was also measured in unpolarized
$pp$, $pN$ fixed-target experiments, see e.g. Ref.~\cite{Heller:1996pg}. Also in this case
it will be hopefully possible in the near future
to check if this intriguing effect survives at the much
larger energies reachable at RHIC and at the Large Hadron Collider (LHC) at CERN.
Similar effects, leading to azimuthal asymmetries both in the polarized
and unpolarized case, have been measured in Drell-Yan (DY) processes~\cite{Zhu:2006gx,Zhu:2008sj},
in semi-inclusive deeply inelastic scattering
(SIDIS)~\cite{Airapetian:2009ti,Pappalardo:2008zz,Alekseev:2008dn,Alekseev:2010rw},
 and in hadron-pair production in $e^+e^-$ collisions~\cite{Abe:2005zx,Seidl:2008xc}.

These results cannot be explained at leading-twist (LT) approximation
in the usual collinear approach of perturbative QCD (pQCD),
based on factorization theorems, to inclusive particle production
in hadronic collisions. Here collinear means that intrinsic parton motion is
neglected in the hard scattering processes
and integrated over up to the large energy scale
in the soft functions involved.
On the contrary, at least in the kinematical regimes under
consideration at RHIC, collinear next-to-leading order (NLO)
pQCD gives a fair account of unpolarized cross
sections (see e.g.~Refs.~\cite{Abelev:2009pb,Adler:2003pb}).

Two different main theoretical approaches have been proposed
in the framework of perturbative QCD in order to account
for these measurements.
One is the so-called twist-three collinear approach, which
generalizes the leading order (LO) collinear framework with
the inclusion of higher-twist quark-gluon
correlations~\cite{Efremov:1984ip,Qiu:1991pp,Qiu:1998ia}.
This involves a new class of universal nonperturbative twist-three
quark-gluon distribution and fragmentation functions that need to be
modeled by fitting experimental data.
Another formalism, which will be adopted in this paper,
is the so-called transverse momentum dependent
(TMD) approach, which takes into
account spin and intrinsic parton motion effects.

Although the large single spin asymmetries (SSAs) of interest here
were originally observed in single inclusive particle production in hadronic
collisions, it is now clear that from the theoretical
point of view these are not the cleanest processes to consider.
First of all, these SSAs are twist-three effects in a series expansion
in inverse powers of the large energy scale (here, the transverse momentum of
the observed single hadron or jet). Several competing mechanisms can therefore
play a role and mix up. Moreover, in the TMD formalism factorization has not yet
been proved for these processes and its validity
is much under debate presently, see e.g.~Refs.~\cite{Boer:2003cm,%
Collins:2004nx,Collins:2007jp,Collins:2007nk,Bomhof:2007xt,Rogers:2010dm}
for inclusive two-hadron production.
Factorization breaking would in turn
imply non universality of the soft TMD distribution
and fragmentation functions required to explain data.
In the case of single particle production in hadronic collisions,
therefore, the TMD approach can be seen at present as a useful generalization
of the parton model where factorization is assumed as a reasonable starting point
to be carefully scrutinized by comparison with available experimental results.

As mentioned above, similar spin and azimuthal asymmetries
were subsequently observed in SIDIS and DY processes,
where two energy scales are involved, the required large energy
scale (the momentum transfer $Q$ in SIDIS,
the lepton-pair invariant mass, $M$, in DY) allowing use of pQCD,
and a small energy scale sensitive to intrinsic parton
motion (the transverse momentum respectively of the final lepton pair in DY
and of the produced hadron in SIDIS).
{}For these classes of processes both the twist-three collinear formalism and
a full gauge invariant TMD approach have been developed and factorization
has been proven~\cite{Ji:2002aa,Belitsky:2002sm,Ji:2004wu,Ji:2004xq}.
Moreover, it has been shown that when the value of the
small observed scale is intermediate between the typical QCD nonperturbative
scale and the large factorization scale, the two approaches
are mutually consistent~\cite{Ji:2006ub,Ji:2006vf,Ji:2006br}.

In the TMD approach to DY(SIDIS) processes color gauge invariance
is ensured by the inclusion of gauge links (Wilson lines),
accounting for initial(final) state interactions among the
struck partons involved in the hard process and
the remnants of the parent hadrons
(additional final state interactions are also
present in the fragmentation process).
Single-spin and azimuthal asymmetries are generated
by TMD polarized partonic distribution and fragmentation functions, among which
the most relevant from a phenomenological point of view are the
Sivers distribution~\cite{Sivers:1989cc,Sivers:1990fh}
and, for transversely polarized quarks,
the Boer-Mulders distribution~\cite{Boer:1997nt}
and the Collins fragmentation function~\cite{Collins:1992kk}
(similar functions can be defined for linearly polarized gluons,
see e.g.~Ref.~\cite{Anselmino:2005sh}).

For inclusive forward pion production the large transverse single spin
asymmetry observed can be generated both by the Sivers and the Collins
effects; unfortunately these contributions cannot be disentangled and one
has to consider alternative measurements in order to separate the different
mechanisms. This is at variance with the case of SIDIS and DY processes, where the
Sivers and Collins effects (and several other possible contributions
to the azimuthal asymmetries) can be disentangled.
In hadronic collisions one has to resort to
different processes, like e.g.~the DY process (no fragmentation),
single photon or jet production, two particle(jet) production with
transverse momentum imbalance and so on.

{}From this point of view, a very interesting process is
$pp\to{\rm jet}+\pi+X$, where one observes a large $p_T$ jet
and looks for the azimuthal distribution of leading pions inside the jet.
In this case, one should observe a symmetric pion distribution for the
fragmentation of an unpolarized parton jet, and a $\cos\phi$ ($\cos2\phi$),
distribution for a transversely(linearly) polarized quark(gluon) parton jet
($\phi$ indicates the azimuthal angle of the leading pion distribution
around the jet axis).
Therefore, despite the complexity of the measurement (which is
however at reach and presently under active investigation at RHIC),
this process might offer plenty of new information as compared to
single  inclusive pion production.
It would in principle allow us to disentangle the
contributions coming from the Sivers and the Collins effects.
Other contributions involving different combinations of TMD
distribution and fragmentation functions could also be disentangled.
Finally, it could also help in identifying jets
coming from quark or gluon fragmentation, since
the pion azimuthal distribution is different in the two cases.
At RHIC kinematics a careful tuning of the kinematical
configuration considered can help from this point of view.

Motivated by these considerations,
in this paper we will present, in the approach of the
TMD generalized parton model, and allowing for intrinsic parton
motion both in the initial colliding hadrons and in the
fragmentation process (which is crucial), the general
expression for the polarized cross section for the process
$p^\uparrow p\to{\rm jet}+\pi+X$, and the structure of the
azimuthal asymmetries that can be measured in the distribution
of leading pions around the jet thrust axis (coinciding in our scheme
with the final scattered parton direction of motion).
A very preliminary version of this study was first presented
in Ref.~\cite{dalesio:2007ec}.
A similar analysis was discussed in Ref.~\cite{Yuan:2007nd}, which
however considered intrinsic parton motion only in the fragmentation
process, drastically reducing the possible contributions to the
asymmetry. Indeed, in that case, only the Collins effect for
quarks is at work. In fact, Ref.~\cite{Yuan:2007nd} aimed at
studying only the Collins fragmentation function (FF),
which should be universal, in
a more simplified theoretical scheme
for which factorization has been proven.
Our approach is different in some respects. It is more general
and has in principle a richer structure in the observable
azimuthal asymmetries, since intrinsic motion is taken into
account in the initial hadrons also.
However, since factorization has not been proven in this case, but
is rather taken as a reasonable phenomenological assumption,
the validity of the scheme and the universality of the TMD
distributions involved require an even more severe scrutiny
by comparison with experimental results.
On the other hand, at the present theoretical and
experimental stage, we believe that combined phenomenological
tests of different approaches are required to clarify the
validity of factorization and, related to this, the relevance
of possible universality-breaking terms for the TMD distributions.

The plan of the paper is the following. In Sect.~\ref{sec-formalism} we
will summarize the TMD generalized parton model approach, which has been
presented and discussed at length in a series of papers, see
e.g.~Refs.~\cite{D'Alesio:2004up,Anselmino:2004ky,Anselmino:2005sh}.
We will then present the expression of the polarized cross section for the
process of interest, discussing in detail the different partonic contributions
to the process; we will finally list the azimuthal asymmetries that can be
measured and their physical content.
In Sect.~\ref{sec-results} we will present phenomenological
results for the azimuthal asymmetries discussed in the
kinematical configuration of the RHIC experiments, at
different c.m.~energies and for central and forward
rapidity jet production. In particular,
we will first present results for the totally maximized
effects, by taking all TMD functions saturated to natural
positivity bounds and adding in sign all possible partonic
contributions. This will assess the potential phenomenological
relevance of each effect. We will then consider more
carefully those effects involving the Sivers and
Boer-Mulders distributions and the Collins fragmentation function,
for which phenomenological parameterizations obtained by
fitting combined data for azimuthal asymmetries in SIDIS, Drell-Yan and
$e^+e^-$ collisions are available.
Section~\ref{sec-conclusions} contains our final remarks and
conclusions.
\section{\label{sec-formalism} Formalism}
In this section we present and summarize the expressions of the polarized
cross section and of the measurable azimuthal asymmetries for the process $A^\uparrow B\to {\rm jet}+\pi+X$,
where $A$ and $B$ are typically a $pp$ or $p\bar p$ pair. Since most of the formalism has been
already presented in Refs.~\cite{D'Alesio:2004up,Anselmino:2004ky,Anselmino:2005sh}, we will shortly
recall the main ingredients of the approach, discussing more extensively only relevant details
specific to the process considered.

Within a generalized TMD parton model approach including spin and intrinsic
parton motion effects, and assuming factorization,
the invariant differential cross section for the process $A(S_A)\, B\to {\rm jet}+\pi+X$
can be written, at leading twist in the soft TMD functions, as follows:
\begin{eqnarray}
\frac{E_{\rm j}\,\de\sigma^{A(S_A)\, B\to {\rm jet}+\pi+X}}
{\de^3\bm{p}_{\rm j}\,\de z\,\de^2\bm{k}_{\perp\pi}} &=&
\sum_{a,b,c,d,\{\lambda\}} \int \frac{\de x_a \de x_b}{16\pi^2 x_a x_b s}\,
\de^2\bm{k}_{\perp a}\de^2\bm{k}_{\perp b}\,
\rho^{a/A,S_A}_{\lambda^{}_a\lambda^{\prime}_a}\hat{f}_{a/A,S_A}(x_a,\bm{k}_{\perp a})
\rho^{b/B}_{\lambda^{}_b\lambda^{\prime}_b}\hat{f}_{b/B}(x_b,\bm{k}_{\perp b})\nonumber\\
&\times& \hat{M}_{\lambda^{}_c,\lambda^{}_d;\lambda^{}_a,\lambda^{}_b}
\hat{M}^*_{\lambda^{\prime}_c,\lambda^{}_d;\lambda^{\prime}_a,\lambda^{\prime}_b}
\delta(\hat{s}+\hat{t}+\hat{u})\hat{D}^\pi_{\lambda^{}_c,\lambda^{\prime}_c}(z,\bm{k}_{\perp\pi})\,.
\label{ds-pi-jet}
\end{eqnarray}

In a LO pQCD approach the scattered parton $c$ in the hard
elementary process $ab\to cd$ is identified with the observed fragmentation jet.
Let us summarize briefly the physical meaning of the terms in Eq.~(\ref{ds-pi-jet}).
Full details and technical aspects can be found in Refs.~\cite{D'Alesio:2004up,Anselmino:2004ky,Anselmino:2005sh}.

We sum over all allowed partonic processes contributing to the physical process observed.
$\{\lambda\}$ stays for a sum over all partonic helicities, $\lambda = \pm 1/2\,(\pm 1)$ for quark(gluon)
partons respectively. $x_{a,b}$ and $\bm{k}_{\perp a,b}$ are respectively the initial parton light-cone
momentum fractions and intrinsic transverse momenta. Analogously, $z$ and $\bm{k}_{\perp\pi}$ are
the light-cone momentum fraction and the transverse momentum of the observed pion inside the jet
with respect to (w.r.t.) the jet (parton $c$) direction of motion.

$\rho^{a/A,S_A}_{\lambda^{}_a\lambda^{\prime}_a}\hat{f}_{a/A,S_A}(x_a,\bm{k}_{\perp a})$
contains all information on the polarization state of the initial parton $a$, which depends
in turn on the (experimentally fixed) parent hadron $A$ polarization state and on the
soft, nonperturbative dynamics encoded in the eight leading-twist polarized and transverse
momentum dependent parton distribution functions, which will be discussed in the following. $\rho^{a/A,S_A}_{\lambda^{}_a\lambda^{\prime}_a}$ is the helicity
density matrix of parton $a$.
Analogously, the polarization state of parton $b$ inside the unpolarized
hadron $B$ is encoded into
$\rho^{b/B}_{\lambda^{}_b\lambda^{\prime}_b}\hat{f}_{b/B}(x_b,\bm{k}_{\perp b})$.

The $\hat{M}_{\lambda^{}_c,\lambda^{}_d;\lambda^{}_a,\lambda^{}_b}$'s are the
pQCD leading-order helicity scattering amplitudes for the hard partonic process $ab\to cd$.

The $\hat{D}^\pi_{\lambda^{}_c,\lambda^{\prime}_c}(z,\bm{k}_{\perp\pi})$'s are the soft
leading-twist TMD fragmentation functions describing the fragmentation process of the
scattered (polarized) parton $c$ into the final leading pion inside the jet.

As already said, we will consider as initial particles $A$, $B$, two spin $1/2$ hadrons
(typically, two protons) with hadron $B$ unpolarized and hadron $A$ in a pure transverse
spin state denoted by $S_A$, with polarization (pseudo)vector $\bm{P}^A$.

$E_{\rm j}$ and $\bm{p}_{\rm j}$ are respectively the energy
and three-momentum of the observed jet.

Unless otherwise stated, we will
always work in the $AB$ hadronic c.m.~frame, with hadron $A$ moving
along the $+\hat{\bm{Z}}_{\rm cm}$ direction; we will define $(XZ)_{\rm cm}$ as the
production plane containing the colliding beams and the observed jet,
with $(\bm{p}_{\rm j})_{X_{\rm cm}}>0$. We therefore have, neglecting
all masses (see also Fig.~\ref{fig-kinem}):
\begin{figure*}[t]
 \includegraphics[angle=0,width=0.8\textwidth]{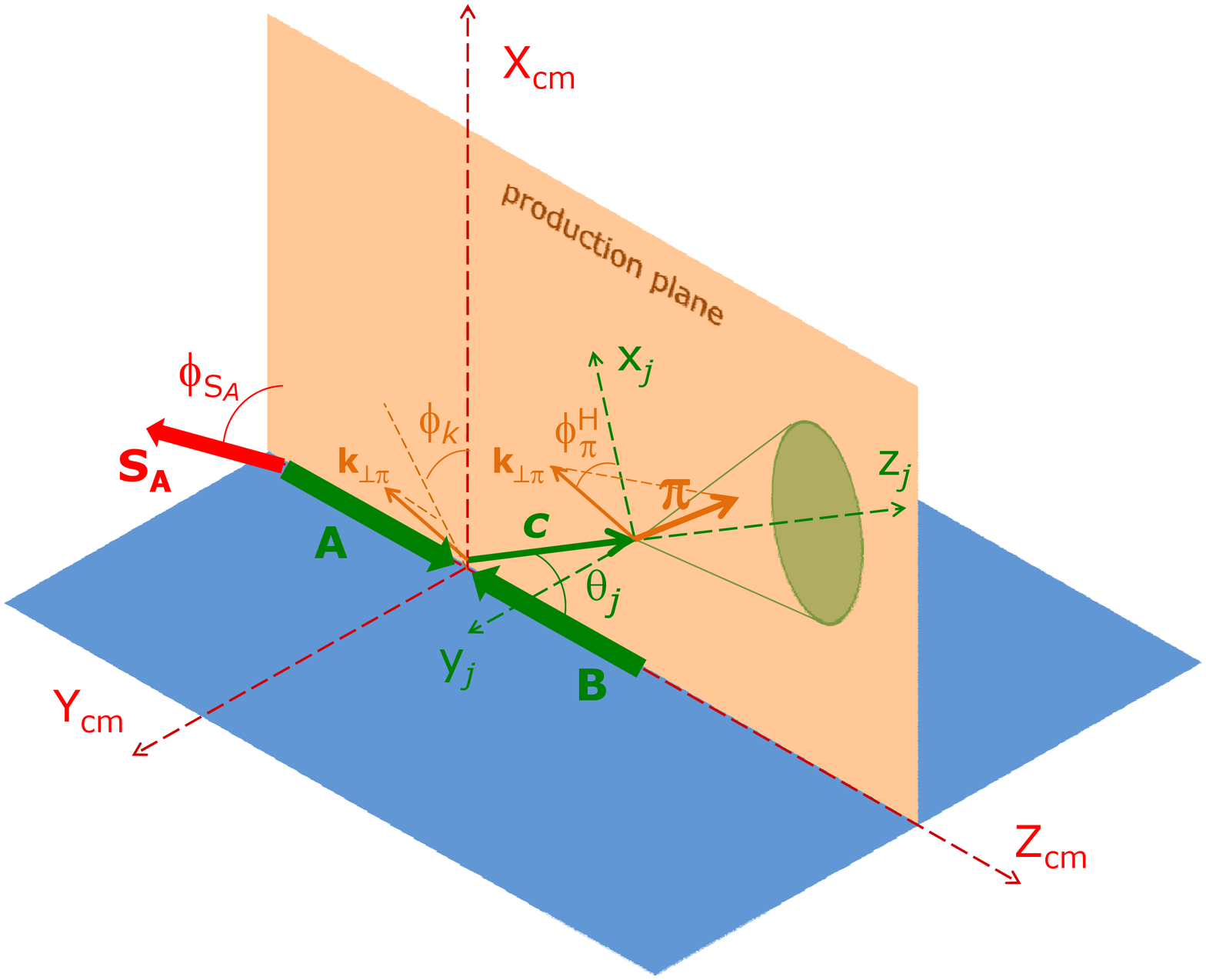}
 \caption{(color online). Kinematical configuration for the
 process $A(S_A)B\to {\rm jet}+\pi+X$ in the
 hadronic c.m.~reference frame.
 \label{fig-kinem} }
\end{figure*}
\begin{eqnarray}
p_A^\mu &=& \frac{\sqrt{s}}{2}(1,0,0,1) \nonumber\\
S_A^\mu &=& S_T^\mu = (0,\cos\phi_{S_A},\sin\phi_{S_A},0)\nonumber\\
p_B^\mu &=& \frac{\sqrt{s}}{2}(1,0,0,-1) \nonumber\\
p_a^\mu &=& \left(x_a\frac{\sqrt{s}}{2}+\frac{\bm{k}_{\perp a}^2}{2x_a\sqrt{s}},
k_{\perp a}\cos\phi_a,k_{\perp a}\sin\phi_a,
x_a\frac{\sqrt{s}}{2}-\frac{\bm{k}_{\perp a}^2}{2x_a\sqrt{s}}\right)\nonumber\\
p_b^\mu &=& \left(x_b\frac{\sqrt{s}}{2}+\frac{\bm{k}_{\perp b}^2}{2x_b\sqrt{s}},
k_{\perp b}\cos\phi_b,k_{\perp b}\sin\phi_b,
-x_b\frac{\sqrt{s}}{2}+\frac{\bm{k}_{\perp b}^2}{2x_b\sqrt{s}}\right)\nonumber\\
p_c^\mu &\equiv& p_{\rm j}^\mu = (E_{\rm j}, p_{{\rm j}\,T},0,p_{{\rm j}\,L}) =
E_{\rm j}(1,\sin\theta_{\rm j},0,\cos\theta_{\rm j}) =
p_{{\rm j}\,T}(\cosh \eta_{\rm j},1,0,\sinh \eta_{\rm j}) \nonumber\\
p_{\pi}^\mu &=&
E_{\pi}(1,\sin\theta_{\pi}\cos\phi_\pi,\sin\theta_{\pi}\sin\phi_\pi,\cos\theta_{\pi})\,,
\label{4mom-cm}
\end{eqnarray}
where $\eta_{\rm j}$ is the jet (pseudo)rapidity, $\eta_{\rm j} = -\log[\tan(\theta_{\rm j}/2)]$.

Notice that, since the observed jet is identified with the scattered parton $c$,
the helicity frame of the fragmenting parton, whose $z$ axis, $\hat{\bm{z}}_{\rm j}$, is along the
direction of motion of parton $c$, is related to the hadronic c.m.~frame by a simple rotation
by $\theta_{\rm j}$ around $\hat{\bm{Y}}_{\rm cm}\equiv \hat{\bm{y}}_{\rm j}$.
In this frame ($\hat{\bm{z}}_{\rm j}$ identifies also the jet
light-cone direction) we have:
\begin{eqnarray}
\tilde{p}_c^\mu &=& \tilde{p}_{\rm j}^\mu = E_{\rm j}(1,0,0,1)\nonumber\\
\tilde{p}_\pi^\mu &=& \left(E_\pi,\bm{k}_{\perp\pi},\sqrt{E_\pi^2-\bm{k}_{\perp\pi}^2}\,\right) =
\left(E_\pi,{k}_{\perp\pi}\cos\phi_\pi^H,{k}_{\perp\pi}
\sin\phi_\pi^H,\sqrt{E_\pi^2-\bm{k}_{\perp\pi}^2}\,\right)\,,
\label{4mom-H}
\end{eqnarray}
where $\phi_\pi^H$ is the azimuthal
angle of the pion three-momentum around the jet direction of motion, as measured in the
fragmenting parton helicity frame.

The light-cone momentum fraction of the observed pion is given by
\begin{equation}
z = \frac{p_\pi^+}{p_c^+}\equiv \frac{p_\pi^+}{p_{\rm j}^+} =
\frac{E_\pi+\sqrt{E_\pi^2-\bm{k}_{\perp\pi}^2}}{2E_{\rm j}}\, .
\label{z-def}
\end{equation}

We can also write, respectively in the jet (parton $c$) helicity frame and in the hadronic
c.m.~frame:
\begin{eqnarray}
\bm{p}_\pi &=& k_{\perp\pi}\cos\phi_\pi^H\hat{\bm{x}}_{\rm j}+
k_{\perp\pi}\sin\phi_\pi^H\hat{\bm{y}}_{\rm j}+
\sqrt{E_\pi^2-\bm{k}_{\perp\pi}^2}\,\hat{\bm{z}}_{\rm j}\nonumber\\
&=& \Bigl[\,k_{\perp\pi}\cos\phi_\pi^H\cos\theta_{\rm j}+
\sqrt{E_\pi^2-\bm{k}_{\perp\pi}^2}\sin\theta_{\rm j}\,\Bigr]\hat{\bm{X}}_{\rm cm}+
k_{\perp\pi}\sin\phi_\pi^H\hat{\bm{Y}}_{\rm cm} \label{pi-H-cm}\\
&+&\Bigl[\,-k_{\perp\pi}\cos\phi_\pi^H\sin\theta_{\rm j}+
\sqrt{E_\pi^2-\bm{k}_{\perp\pi}^2}\cos\theta_{\rm j}\,\Bigr]\hat{\bm{Z}}_{\rm cm}\,.\nonumber
\end{eqnarray}

Let us stress that in our notation intrinsic transverse momenta, $\bm{k}_{\perp i}$, $i=a,b,\pi$,
are always three-vectors and $k_{\perp i} \equiv |\bm{k}_{\perp i}|$. This has to be kept in
mind when comparing with literature, where often intrinsic momenta are intended as 4-vectors
and $k_{\perp}^2=-\bm{k}_{\perp}^2$.
{}From Eq.~(\ref{pi-H-cm}) it is easy to see that the pion intrinsic transverse momentum is given, in the hadronic
c.m.~frame, by
\begin{equation}
\bm{k}_{\perp\pi} = k_{\perp\pi}\cos\phi_\pi^H\cos\theta_{\rm j}\hat{\bm{X}}_{\rm cm}+
k_{\perp\pi}\sin\phi_\pi^H\hat{\bm{Y}}_{\rm cm}-
k_{\perp\pi}\cos\phi_\pi^H\sin\theta_{\rm j}\hat{\bm{Z}}_{\rm cm}\,.
\label{kpi-cm}
\end{equation}
Defining by $\phi_{k}$ the azimuthal angle of the pion intrinsic transverse momentum,
$\bm{k}_{\perp\pi}$,
\emph{as measured in the hadronic c.m.~frame}, see Fig.~\ref{fig-kinem},
{}from Eq.~(\ref{kpi-cm}) we easily see that
\begin{equation}
\tan\phi_{k} = \frac{\tan\phi_\pi^H}{\cos\theta_{\rm j}}\,.
\label{tan-phik}
\end{equation}

Notice that, for central-rapidity jets ($\theta_{\rm j}=\pi/2$),
$\phi_k=\pi/2$. Therefore, azimuthal asymmetries modulated in terms of $\phi_k$
are artificially suppressed in the central rapidity region, while the physically relevant
angle is $\phi_\pi^H$. Instead, in the forward rapidity region, when $\cos\theta_{\rm j}\to 1$,
the two angles are practically coincident.
Notice also that in Ref.~\cite{Yuan:2007nd}, where only the
forward rapidity region was considered, the angle
$\phi_k$ (called $\phi_h$ there) was adopted.

Let us now come back to the soft TMD partonic distribution and fragmentation functions entering
the differential cross section for the process $A(S_A)\,B\to {\rm jet}+\pi+X$, Eq.~(\ref{ds-pi-jet}).
Consider first the polarized soft process at the distribution level, $A(S_A)\to a +X$;
as said, in Eq.~(\ref{ds-pi-jet}) all information on this process is encoded in the
factor
$\rho^{a/A,S_A}_{\lambda^{}_a\lambda^{\prime}_a}\hat{f}_{a/A,S_A}(x_a,\bm{k}_{\perp a})$.
This factor depends on the polarization state
(fixed by experimental conditions)
of the parent hadron $A$,
described by its own helicity density
matrix $\rho^{A,S_A}$, and on generalized soft distribution functions
for the process $A(S_A)\to a +X$,
$\hat{F}_{\lambda^{}_A,\lambda^{\prime}_A}^{\lambda^{}_a,\lambda^{\prime}_a}(x_a,\bm{k}_{\perp a})$:
\begin{equation}
\rho^{a/A,S_A}_{\lambda^{}_a\lambda^{\prime}_a}\hat{f}_{a/A,S_A}(x_a,\bm{k}_{\perp a})=
\sum_{\lambda^{}_A,\lambda^{\prime}_A}
\rho^{A,S_A}_{\lambda^{}_A,\lambda^{\prime}_A}
\hat{F}_{\lambda^{}_A,\lambda^{\prime}_A}^{\lambda^{}_a,\lambda^{\prime}_a}(x_a,\bm{k}_{\perp a})\,.
\label{rhA-F}
\end{equation}
The functions $\hat{F}_{\lambda^{}_A,\lambda^{\prime}_A}^{\lambda^{}_a,\lambda^{\prime}_a}$
are related to the well-known leading-twist hand-bag diagram for deeply inelastic
scattering.
Analogous relations hold for parton $b$ inside the (un)polarized hadron $B$.

Rotational invariance and parity conservation for strong interactions imply some very general
relations for these nonperturbative functions:
\begin{equation}
\hat{F}_{\lambda^{}_A,\lambda^{\prime}_A}^{\lambda^{}_a,\lambda^{\prime}_a}(x_a,\bm{k}_{\perp a})=
F_{\lambda^{}_A,\lambda^{\prime}_A}^{\lambda^{}_a,\lambda^{\prime}_a}(x_a,k_{\perp a})\,
e^{i(\lambda^{}_A-\lambda^{\prime}_A)\phi_a}\,,
\label{F-rot}
\end{equation}
where $\phi_a$ is the azimuthal angle of parton $a$ intrinsic transverse momentum, $\bm{k}_{\perp a}$,
in the parent hadron $A$ helicity frame (coinciding in the present case with the hadronic c.m.~frame).
Notice that the \emph{reduced} soft functions $F$ on the right-hand side (without the hat)
do not depend anymore on azimuthal phases.
Moreover,
\begin{equation}
F_{-\lambda^{}_A,-\lambda^{\prime}_A}^{-\lambda^{}_a,-\lambda^{\prime}_a}(x_a,k_{\perp a})=
(-1)^{(S_A-s_a)}(-1)^{(\lambda^{}_A-\lambda^{}_a)+(\lambda^{\prime}_A-\lambda^{\prime}_a)}
F_{\lambda^{}_A,\lambda^{\prime}_A}^{\lambda^{}_a,\lambda^{\prime}_a}(x_a,k_{\perp a})\,,
\label{F-par}
\end{equation}
where $S_A$ and $s_a$ are the spins of the parent hadron and of the parton respectively.
Notice that for spin 1/2 colliding hadrons, $S_A=1/2$, the factor $(-1)^{(S_A-s_a)}$ is positive(negative)
for quark(gluon) partons, and therefore some parity properties of the soft functions
are different for quarks and gluons. By combining complex conjugation and parity properties,
one can show that some of these functions are purely real or purely imaginary.
As a result, for spin 1/2 hadrons only eight independent soft distributions survive at leading twist.
They can be easily related to the TMD distributions widely discussed in the literature.
Before listing them for completeness, let us recall another important property of the hadronic
functions ${F}_{\lambda^{}_A\lambda^{\prime}_A}^{\lambda^{}_a\lambda^{\prime}_a}(x_a,{k}_{\perp a})$,
coming from total angular momentum conservation in the forward direction, that is for $k_{\perp a}\to 0$:
\begin{equation}
F_{\lambda^{}_A,\lambda^{\prime}_A}^{\lambda^{}_a,\lambda^{\prime}_a}(x_a,k_{\perp a}) \sim
\left(\frac{k_{\perp a}}{M}\right)^{|\lambda^{}_A-\lambda^{}_a-(\lambda^{\prime}_A-\lambda^{\prime}_a)|}\,
\tilde{F}_{\lambda^{}_A,\lambda^{\prime}_A}^{\lambda^{}_a,\lambda^{\prime}_a}(x_a,k_{\perp a})\,,
\label{kt-power}
\end{equation}
where $M$ is a typical hadronic mass scale and the $\tilde{F}$'s
stay for the remaining part of the functions
that, depending on the details of dynamics, may or may not vanish
in the collinear configuration. The relevant hadronic functions and their
connection with the leading-twist TMD distributions is therefore, in the quark case
(for clarity we adopt here both the notation of Ref.~\cite{Anselmino:2005sh} and that of the
Amsterdam group~\cite{Mulders:1995dh,Boer:1997nt,Boer:2003cm})
\begin{eqnarray}
F_{\!\!q\,++}^{\,++}(x,k_\perp) + F_{\!\!q\,++}^{\,--}(x,k_\perp) &=&
 f_{q/A}(x,k_\perp) = f_{1}^q(x,k_\perp) \nonumber\\
F_{\!\!q\,++}^{\,++}(x,k_\perp) - F_{\!\!q\,++}^{\,--}(x,k_\perp) &=&
 \Delta_L f_{q/A}(x,k_\perp) = g^q_{1L}(x,k_\perp) \nonumber\\
F_{\!\!q\,+-}^{\,+-}(x,k_\perp) &=& h_{1}^q(x,k_\perp) \nonumber\\
F_{\!\!q\,+-}^{\,-+}(x,k_\perp) &=& \frac{k_\perp^2}{2M^2}\,h_{1T}^{\perp q}(x,k_\perp)\nonumber\\
{\rm Re}F_{\!\!q\,+-}^{\,++}(x,k_\perp) &=& \frac{k_\perp}{2M}\,g_{1T}^{\perp q}(x,k_\perp)\nonumber\\
{\rm Im}F_{\!\!q\,+-}^{\,++}(x,k_\perp) &=& \frac{1}{4}\,\Delta^{\!N}f_{q/A^\uparrow}(x,k_\perp) =
-\frac{k_\perp}{2M}\,f_{1T}^{\perp q}(x,k_\perp)\nonumber\\
{\rm Re}F_{\!\!q\,++}^{\,+-}(x,k_\perp) &=& \frac{k_\perp}{2M}\,h_{1L}^{\perp q}(x,k_\perp)\nonumber\\
{\rm Im}F_{\!\!q\,++}^{\,+-}(x,k_\perp) &=& -\frac{1}{2}\,\Delta^{\!N}f_{q^\uparrow/A}(x,k_\perp) =
\frac{k_\perp}{2M}\,h_{1}^{\perp q}(x,k_\perp)\,.
\label{F-notation}
\end{eqnarray}

Analogous relations hold for gluon partons with the changes discussed above, due
to the different spin of the parton and leading to different parity properties.
Notice that instead of transversely polarized quarks we will have linearly polarized gluons.
Of course, the same relations hold also for the $B\to b+X$ process. However,
this time the $\hat{\bm{X}}_B$ and $\hat{\bm{Z}}_B$ axes of hadron $B$ helicity frame
are opposite to those of the hadronic c.m.~frame.

Concerning the fragmentation process, since here we are considering only pions
(in general, unpolarized hadrons), the discussion
of the soft fragmentation functions, $\hat{D}^\pi_{\lambda^{}_c,\lambda^{\prime}_c}$, is much
simplified. In practice, only two independent TMD fragmentation functions survive:
one with diagonal parton helicity indexes, related to the TMD unpolarized FF,
\begin{equation}
\hat{D}^{\pi/c}_{\pm\pm}(z,\bm{k}_{\perp\pi}) \equiv  {D}^{\pi/c}_{\pm\pm}(z,k_{\perp\pi})
= D_{\pi/c}(z,k_{\perp\pi})\,,
\label{d++}
\end{equation}
and a second one with off-diagonal parton helicity indexes, $\hat{D}^{\pi/c}_{+-}(z,\bm{k}_{\perp\pi})$.
This second function is purely imaginary for quark partons and is related to the well-known
Collins function~\cite{Collins:1992kk}, describing the fragmentation of
a transversely polarized quark into a non collinear unpolarized hadron.
Instead, for gluon partons the analogous Collins-like function is purely real
and is related to the fragmentation
of linearly polarized gluons again into an unpolarized hadron.

It is very important to realize that these off-diagonal quark and gluon TMD FFs have
different behaviours as a function of the azimuthal angle of the observed pion
around the direction of motion of the fragmentation jet. More specifically,
for quarks we have:
\begin{equation}
\hat{D}^{\pi/q}_{\pm\mp}(z,\bm{k}_{\perp\pi}) = \pm\, D^{\pi/q}_{+-}(z,{k}_{\perp\pi})e^{\pm i\phi_\pi^H}\,,
\label{off-diag-FF-q}
\end{equation}
while for gluons the analogous relation reads
\begin{equation}
\hat{D}^{\pi/g}_{\pm\mp}(z,\bm{k}_{\perp\pi}) =  D^{\pi/g}_{+-}(z,{k}_{\perp\pi})e^{\pm i 2\phi_\pi^H}\,.
\label{off-diag-FF-g}
\end{equation}
As mentioned above, the azimuthal independent parts of these TMD FFs are related to the probability for a
transversely(linearly) polarized quark(gluon) of fragmenting into a non collinear unpolarized hadron,
the Collins(Collins-like for gluons) fragmentation function~\cite{Anselmino:2005sh},
\begin{eqnarray}
D^{\pi/q}_{+-}(z,{k}_{\perp\pi}) &=& \frac{i}{2}\,\Delta^{\!N}D_{\pi/q^\uparrow}(z,{k}_{\perp\pi})
= \frac{i}{2}\frac{k_{\perp\pi}}{zm_\pi}\,H_1^{\perp}(z,{k}_{\perp\pi})\nonumber\\
D^{\pi/g}_{+-}(z,{k}_{\perp\pi}) &=&
\frac{1}{2}\,\Delta^{\!N}D_{\pi/{\cal T}_1^g}(z,{k}_{\perp\pi})\,.
\label{Coll-q-g}
\end{eqnarray}

Again, total angular momentum conservation in the forward direction dictates the
power behaviour of the TMD FFs for $k_{\perp\pi}\to 0$:
\begin{equation}
{D}^\pi_{\lambda^{}_c,\lambda^{\prime}_c}(z,k_{\perp\pi}) \sim
\left(\frac{k_{\perp\pi}}{M}\right)^{|\lambda^{}_c-\lambda^{\prime}_c|}\,
\tilde{D}^\pi_{\lambda^{}_c,\lambda^{\prime}_c}(z,k_{\perp\pi})\,.
\label{kt-power-D}
\end{equation}
Also in this case, the behaviour of the off-diagonal FFs is different for quarks and
gluons.

Let us finally comment on the helicity amplitudes for the partonic hard scattering
processes entering Eq.~(\ref{ds-pi-jet}). Again, details have been already presented in
Refs.~\cite{Anselmino:2004ky,Anselmino:2005sh} and we limit here to summarize some
useful properties.
Due to intrinsic partonic motion in the distributions and in the
fragmentation process, the general kinematical configuration for the partonic process
$ab\to cd$ is not planar in the hadronic c.m.~frame. Since partons can be in general
polarized in the process, azimuthal phases are therefore essential and must be properly taken into account.
For massless partons, due to helicity conservation in the quark-gluon vertex and
parity invariance only three independent helicity amplitudes survive:
\begin{eqnarray}
\hat{M}_{++;++} &=& \hat{M}^*_{--;--} = \hat{M}_1^0 e^{i\varphi_1}\nonumber\\
\hat{M}_{-+;-+} &=& \hat{M}^*_{+-;+-} = \hat{M}_2^0 e^{i\varphi_2}\\
\hat{M}_{-+;+-} &=& \hat{M}^*_{+-;-+} = \hat{M}_3^0 e^{i\varphi_3}\,.\nonumber
\label{hel-ampl}
\end{eqnarray}
Here $\pm$ stays for $\lambda=\pm 1/2$ for quarks and $\lambda=\pm 1$ for gluons.
$\hat{M}^0_i$ ($i=1,2,3$) are the three independent helicity amplitudes in the
canonical partonic c.m.~frame, that is a frame where partons $a$, $b$ move along the
$\pm\hat{\bm{z}}$ direction respectively and the scattering plane coincides with the $(xz)$ plane.
The phases $\varphi_i$ collect all azimuthal phases coming from rotations and boosts
connecting the canonical partonic c.m.~frame with the hadronic c.m.~frame
adopted in the paper. Their general expression is rather involved. All details
and the explicit expressions of the $\hat{M}^0_i$ and $\varphi_i$
can be found in Refs.~\cite{Anselmino:2004ky,Anselmino:2005sh}.
Here we limit ourselves to notice that for parton $c$ lying in the $(XZ)_{\rm cm}$ plane,
as in the present case, the phases $\varphi_i$ are odd under
$\bm{k}_{\perp a,b} \to -\bm{k}_{\perp a,b}$. This property, as we will see below,
is very helpful in selecting physically observable effects out of the many
contributions present in Eq.~(\ref{ds-pi-jet}) because of the nonplanarity of the
partonic process, which however do not survive at the hadronic level under
integration over intrinsic parton momenta.

We now concentrate on the partonic kernels entering the expression of the
polarized cross section, Eq.~(\ref{ds-pi-jet}):
\begin{eqnarray}
\Sigma(S_A)^{ab\to cd} &=& \sum_{\{\lambda\}}
\rho^{a/A,S_A}_{\lambda^{}_a\lambda^{\prime}_a}\hat{f}_{a/A,S_A}(x_a,\bm{k}_{\perp a})
\rho^{b/B}_{\lambda^{}_b\lambda^{\prime}_b}
\hat{f}_{b/B}(x_b,\bm{k}_{\perp b})\nonumber\\
&\times& \hat{M}_{\lambda^{}_c,\lambda^{}_d;\lambda^{}_a,\lambda^{}_b}
\hat{M}^*_{\lambda^{\prime}_c,\lambda^{}_d;\lambda^{\prime}_a,\lambda^{\prime}_b}
\hat{D}^\pi_{\lambda^{}_c,\lambda^{\prime}_c}(z,\bm{k}_{\perp\pi})\,.
\label{gen-kern}
\end{eqnarray}

One has to evaluate the kernels for each of the eight distinct partonic channels
contributing to the cross section,
\begin{eqnarray}
&& qq\to qq,\quad qg\to qg, \quad qg\to gq, \quad gq\to qg, \quad gq\to gq\\\nonumber
&& gg\to q\bar{q}, \quad q\bar{q}\to gg, \quad gg\to gg\,,
\label{channels}
\end{eqnarray}
where in the first line $q$ stays for both quarks and antiquarks in all allowed
combinations.

In practice, the calculation is performed by summing explicitly
over all helicity indexes and inserting the appropriate expressions for the
helicity density matrices of partons $a$, $b$ and for the polarized distribution
and fragmentation functions, as detailed above.
Furthermore, after factorizing explicitly all azimuthal dependences,
including those coming from the hard-scattering helicity amplitudes, collecting
them and using symmetry properties under $\bm{k}_{\perp a,b}\to - \bm{k}_{\perp a,b}$,
one gets the final expression for the kernels, containing only physically allowed
terms at the hadronic level.

 We will not present explicitly the kernels for all channels.
Instead, we limit ourselves to give the kernels for the $qq\to qq$ and $gg\to gg$ channels,
which contain the maximal number of terms and give examples of possible contributions
involving both quark and gluon distribution and fragmentation functions.
Moreover, we will directly present the combination of kernels,
$\Sigma(\phi_{S_A})\pm\Sigma(\phi_{S_A}+\pi)$ entering the numerator and the denominator of
the single spin azimuthal asymmetries discussed in the sequel. We will also omit for shortness
the explicit dependences on the light-cone momentum fractions and intrinsic momenta of all
TMD distribution and fragmentation functions. On the contrary,
all azimuthal dependences are explicitly shown. In particular, terms are collected according
to the azimuthal dependence in the fragmentation process which directly enter
the azimuthal asymmetries we want to study.
Therefore, we get for the $qq\to qq$ channel:
\begin{eqnarray}
&&\bigl[\,\Sigma(\phi_{S_A}) + \Sigma(\phi_{S_A}+\pi)\,\bigr]^{qq\to qq} \nonumber\\
&&\sim\Big\{\,f_{a/A}f_{b/B}\,[\,|\hat{M}^0_1|^2+|\hat{M}^0_2|^2+|\hat{M}^0_3|^2\,]\nonumber\\
&&\quad\quad -2\Delta^{\!N}f_{a^\uparrow/A}\Delta^{\!N}f_{b^\uparrow/B}
\hat{M}^0_2\hat{M}^0_3\cos(\varphi_2-\varphi_3)\,\Big\}D_{\pi/q}\nonumber\\
&&+\,\Big\{\,\Delta^{\!N}f_{a^\uparrow/A}f_{b/B}
\hat{M}^0_1\hat{M}^0_2\cos(\varphi_1-\varphi_2)\nonumber\\
&&\quad\quad-f_{a/A}\Delta^{\!N}f_{b^\uparrow/B}
\hat{M}^0_1\hat{M}^0_3\cos(\varphi_1-\varphi_3)\,\Big\}\Delta^{\!N}D_{\pi/q^\uparrow}
\cos\phi_\pi^H\,.
\label{dsig-unpol-qq}
\end{eqnarray}

The symbol $\sim$ is to recall that, as discussed above,
this expression is valid only after integrating
over the azimuthal angles of the initial intrinsic parton momenta, $\bm{k}_{\perp a,b}$,
and is based on symmetry properties of the kernels under
$\bm{k}_{\perp a,b}\to-\bm{k}_{\perp a,b}$.
It contains less terms than the general expression for the kernels in the
$\bm{k}_{\perp a,b}$-unintegrated, non planar partonic configuration.

Analogously, for the numerator of the asymmetry, we find:
\begin{eqnarray}
&&\bigl[\,\Sigma(\phi_{S_A}) - \Sigma(\phi_{S_A}+\pi)\,\bigr]^{qq\to qq} \nonumber\\
&&\sim\Big\{\,\frac{1}{2}\Delta^{\!N}f_{a/A^\uparrow}f_{b/B}\cos\phi_a
\,[\,|\hat{M}^0_1|^2+|\hat{M}^0_2|^2+|\hat{M}^0_3|^2\,]\nonumber\\
&&\quad\quad-h_1^a\Delta^{\!N}f_{b^\uparrow/B}\cos(\phi_a-\varphi_2+\varphi_3)
2\hat{M}^0_2\hat{M}^0_3\nonumber\\
&&\quad\quad+\frac{k_{\perp a}^2}{2M^2}h_{1T}^{\perp a}
\Delta^{\!N}f_{b^\uparrow/B}\cos(\phi_a+\varphi_2-\varphi_3)
2\hat{M}^0_2\hat{M}^0_3
\,\Big\}\sin\phi_{S_A}D_{\pi/q}\nonumber\\
&&+\,\Big\{-\frac{1}{2}\Delta^{\!N}f_{a/A^\uparrow}\Delta^{\!N}f_{b^\uparrow/B}
\cos\phi_a\cos(\varphi_1-\varphi_3)\hat{M}^0_1\hat{M}^0_3\nonumber\\
&&\quad\quad +h_1^a f_{b/B}\cos(\phi_a+\varphi_1-\varphi_2)\hat{M}^0_1\hat{M}^0_2\nonumber\\
&&\quad\quad -\frac{k_{\perp a}^2}{2M^2}h_{1T}^{\perp a}
f_{b/B}\cos(\phi_a-\varphi_1+\varphi_2)\hat{M}^0_1\hat{M}^0_2
\,\Big\}\sin\phi_{S_A}\Delta^{\!N}D_{\pi/q^\uparrow}\cos\phi_\pi^H\nonumber\\
&&+\,\Big\{-\frac{1}{2}\Delta^{\!N}f_{a/A^\uparrow}\Delta^{\!N}f_{b^\uparrow/B}
\sin\phi_a\sin(\varphi_1-\varphi_3)\hat{M}^0_1\hat{M}^0_3\nonumber\\
&&\quad\quad -h_1^a f_{b/B}\cos(\phi_a+\varphi_1-\varphi_2)\hat{M}^0_1\hat{M}^0_2\nonumber\\
&&\quad\quad -\frac{k_{\perp a}^2}{2M^2}h_{1T}^{\perp a}
f_{b/B}\cos(\phi_a-\varphi_1+\varphi_2)\hat{M}^0_1\hat{M}^0_2
\,\Big\}\cos\phi_{S_A}\Delta^{\!N}D_{\pi/q^\uparrow}\sin\phi_\pi^H\,.
\label{delta-sig-qq}
\end{eqnarray}

Let us discuss the physical content of these results.
Eq.~(\ref{dsig-unpol-qq}) gives the contribution of the $qq\to qq$ channel
to (twice) the unpolarized cross section. It contains two terms azimuthally
symmetric in the fragmentation process: the first is the usual term already present
in the collinear factorization scheme, the second one is the possible
contribution due to the Boer-Mulders effect coming from both initial partons.

The last two terms in Eq.~(\ref{dsig-unpol-qq}) might potentially give rise
to an azimuthal asymmetry in the ${\rm jet}\to\pi+X$ \emph{unpolarized} process: they are
related to the combined action of the Boer-Mulders function
(either for parton quark $a$ or $b$ separately) and of the Collins
fragmentation function. Notice that only the first contribution
in Eq.~(\ref{dsig-unpol-qq}) survives in a
purely collinear scheme, or even in a scheme, like that adopted in
Ref.~\cite{Yuan:2007nd,Yuan:2008yv,Yuan:2008tv}, where intrinsic
motion is kept into account only in the fragmentation process.

Eq.~(\ref{delta-sig-qq}) is related to transverse spin
asymmetries for pion production inside a jet. Again, it contains terms related
to the unpolarized pion fragmentation function $D_{\pi/q}(z,k_{\perp\pi})$,
which are symmetric with respect to $\phi_\pi^H$, and terms
proportional to the Collins fragmentation function which are
responsible for the azimuthal asymmetries in the jet fragmentation process.

The first group of terms, proportional to $D_{\pi/q}$,
are related to a single spin asymmetry (only hadron
$A$ is polarized). The only contribution allowed by rotational invariance
and parity conservation comes from hadron $A$ being polarized transversely to
the jet production plane, which explains the $\sin\phi_{S_A}$ term.

Notice once more that the appearance of only the physically allowed contributions is not
trivial in our expression, which, at this stage, is still unintegrated over
$\bm{k}_{\perp a,b}$. However, as discussed above, taking into account symmetry
properties under $\bm{k}_{\perp a,b}\to -\bm{k}_{\perp a,b}$ amounts
to select from the beginning only physically allowed contributions from the wealth
of partonic terms present in the general non planar configuration for the partonic
process.

In the case of  Eq.~(\ref{delta-sig-qq}) the terms proportional to
$D_{\pi/q}$ come from the Sivers effect (first term) and from combinations
of the transversity and the Boer-Mulders functions for partons $a$ and $b$
respectively.

Let us now consider the terms in Eq.~(\ref{delta-sig-qq}) related to the
Collins fragmentation function, $\Delta^{\!N}D_{\pi/q^\uparrow}$.
These refer effectively to a
double spin asymmetry, since both hadron $A$ and the final quark $c$
(generating the observed jet) are transversely polarized.
Their physical content is also easy to understand.
The first three terms, proportional to
$\Delta^{\!N}D_{\pi/q^\uparrow}\sin\phi_{S_A}$, correspond
to a double transverse spin asymmetry, where both hadron $A$ and the final parton $c$ are
transversely polarized w.r.t.~the jet production plane, along
the $\hat{\bm{Y}}_{\rm cm}$ axis. In this case,
the Collins effect in the fragmentation process survives only
for the component of the pion transverse momentum (w.r.t.~the jet)
orthogonal to the parton $c$ polarization vector,
that is the component laying in the production plane, which
explains the associated $\cos\phi_\pi^H$ factor (see Fig.~\ref{fig-kinem}).

Analogously, the three terms proportional to
$\Delta^{\!N}D_{\pi/q^\uparrow}\cos\phi_{S_A}$ correspond
again to a double transverse spin asymmetry, this time for hadron $A$ and
the final parton $c$ transversely polarized w.r.t.~their own direction of motion but in
the production plane (i.e.~along the $x$ axis of the respective helicity frames).
Again, only the component of the pion transverse momentum orthogonal to the
parton $c$ polarization vector
contributes to the Collins asymmetry in the fragmentation process,
which is this time guaranteed by the $\sin\phi_\pi^H$ factor.

Although Eq.~(\ref{delta-sig-qq}) makes the physical content of the asymmetry more
evident, the following equivalent expression, where the terms
proportional to the Collins FF are collected differently,
makes the possible azimuthal asymmetries in the jet
fragmentation process more readable:
\begin{eqnarray}
&&\bigl[\,\Sigma(\phi_{S_A}) - \Sigma(\phi_{S_A}+\pi)\,\bigr]^{qq\to qq} \nonumber\\
&&\sim \Big\{\,\frac{1}{2}\Delta^{\!N}f_{a/A^\uparrow}f_{b/B}\cos\phi_a
\,[\,|\hat{M}^0_1|^2+|\hat{M}^0_2|^2+|\hat{M}^0_3|^2\,]\nonumber\\
&&\quad\quad -h_1^a\Delta^{\!N}f_{b^\uparrow/B}\cos(\phi_a-\varphi_2+\varphi_3)
2\hat{M}^0_2\hat{M}^0_3\nonumber\\
&&\quad\quad +\frac{k_{\perp a}^2}{2M^2}h_{1T}^{\perp a}
\Delta^{\!N}f_{b^\uparrow/B}\cos(\phi_a+\varphi_2-\varphi_3)
2\hat{M}^0_2\hat{M}^0_3
\,\Big\}\sin\phi_{S_A}D_{\pi/q}\nonumber\\
&&+\,\Big\{\,\Big[\,h_1^a f_{b/B}\cos(\phi_a+\varphi_1-\varphi_2)
\hat{M}^0_1\hat{M}^0_2\nonumber\\
&&\quad\quad -\frac{1}{4}\Delta^{\!N}f_{a/A^\uparrow}\Delta^{\!N}f_{b^\uparrow/B}
\cos(\phi_a+\varphi_1-\varphi_3)\hat{M}^0_1\hat{M}^0_3
\,\Big]\sin(\phi_{S_A}-\phi_\pi^H)\nonumber\\
&&\quad -\Big[\,\frac{k_{\perp a}^2}{2M^2}h_{1T}^{\perp a}
f_{b/B}\cos(\phi_a-\varphi_1+\varphi_2)\hat{M}^0_1\hat{M}^0_2\nonumber\\
&&\quad\quad +\frac{1}{4}\Delta^{\!N}f_{a/A^\uparrow}\Delta^{\!N}f_{b^\uparrow/B}
\cos(\phi_a-\varphi_1+\varphi_3)\hat{M}^0_1\hat{M}^0_3\,\Big]
\sin(\phi_{S_A}+\phi_\pi^H)\,\Big\}
\Delta^{\!N}D_{\pi/q^\uparrow}\,.
\label{delta-sig-qq-2}
\end{eqnarray}

Therefore, apart from the term proportional to $D_{\pi/q}$,
two azimuthal asymmetries in the distribution of leading pions inside
the jet are possible, proportional respectively to $\sin(\phi_{S_A}\mp\phi_\pi^H)$.

Neglecting intrinsic motion of the initial partons,
$\bm{k}_{\perp a,b}\to \bm{0}$, Eqs.~(\ref{dsig-unpol-qq}),
(\ref{delta-sig-qq-2}) simplify considerably:
\begin{equation}
\bigl[\,\Sigma(\phi_{S_A}) + \Sigma(\phi_{S_A}+\pi)\,\bigr]^{qq\to qq} \,\to\,
 f_{a/A}(x_a)f_{b/B}(x_b)\,[\,|\hat{M}^0_1|^2+|\hat{M}^0_2|^2+|\hat{M}^0_3|^2\,]
D_{\pi/q}(z,k_{\perp\pi})\,,
\label{dsig-unpol-qq-col}
\end{equation}
\begin{equation}
\bigl[\,\Sigma(\phi_{S_A}) - \Sigma(\phi_{S_A}+\pi)\,\bigr]^{qq\to qq} \,\to\,
 h_1^a(x_a) f_{b/B}(x_b)\hat{M}^0_1\hat{M}^0_2
\Delta^{\!N}D_{\pi/q^\uparrow}(z,k_{\perp\pi})
\sin(\phi_{S_A}-\phi_\pi^H)\,,
\label{delta-sig-qq-2-col}
\end{equation}
in agreement with the results of Ref.~\cite{Yuan:2007nd}. Notice however that our angle $\phi_\pi^H$
is the azimuthal angle of $\bm{k}_{\perp\pi}$ measured in the jet (parton c) helicity frame.
Therefore, it does not coincide with the angle $\phi_h$ utilized in Ref.~\cite{Yuan:2007nd},
which is the azimuthal angle of $\bm{k}_{\perp\pi}$ measured in the hadronic c.m.~frame
[we call this angle $\phi_k$ in this paper, see Eq.~(\ref{tan-phik})
and Fig.~\ref{fig-kinem}]. Only for forward
jet production ($\cos\theta_{\rm j}\to 1$) do these angles coincide with good approximation.

In the case of the $gg\to gg$ partonic channel, the analogues of Eqs.~(\ref{dsig-unpol-qq}),
(\ref{delta-sig-qq-2}) are:
\begin{eqnarray}
&&\bigl[\,\Sigma(\phi_{S_A}) + \Sigma(\phi_{S_A}+\pi)\,\bigr]^{gg\to gg} \nonumber\\
&&\sim \Big\{\,f_{a/A}f_{b/B}\,[\,|\hat{M}^0_1|^2+|\hat{M}^0_2|^2+|\hat{M}^0_3|^2\,]\nonumber\\
&&\quad\quad -2\Delta^{\!N}f_{{\cal T}_1^a/A}\Delta^{\!N}f_{{\cal T}_1^b/B}
\hat{M}^0_2\hat{M}^0_3\cos(\varphi_2-\varphi_3)\,\Big\}D_{\pi/g}\nonumber\\
&&+\,\Big\{\,\Delta^{\!N}f_{{\cal T}_1^a/A}f_{b/B}
\hat{M}^0_1\hat{M}^0_2\cos(\varphi_1-\varphi_2)\nonumber\\
&&\quad\quad +f_{a/A}\Delta^{\!N}f_{{\cal T}_1^b/B}
\hat{M}^0_1\hat{M}^0_3\cos(\varphi_1-\varphi_3)\,\Big\}\Delta^{\!N}D_{\pi/{\cal T}_1^g}
\cos2\phi_\pi^H\,,
\label{dsig-unpol-gg}
\end{eqnarray}
\begin{eqnarray}
&&\bigl[\,\Sigma(\phi_{S_A}) - \Sigma(\phi_{S_A}+\pi)\,\bigr]^{gg\to gg} \nonumber\\
&&\sim \Big\{\,\frac{1}{2}\Delta^{\!N}f_{a/A^\uparrow}f_{b/B}\cos\phi_a
\,[\,|\hat{M}^0_1|^2+|\hat{M}^0_2|^2+|\hat{M}^0_3|^2\,]\nonumber\\
&&\quad\quad +{\rm Im}F_{\!\!a\,+-}^{\,+-}\Delta^{\!N}f_{{\cal T}_1^b/B}
\cos(\phi_a-\varphi_2+\varphi_3)2\hat{M}^0_2\hat{M}^0_3\nonumber\\
&&\quad\quad +{\rm Im}F_{\!\!a\,+-}^{\,-+}
\Delta^{\!N}f_{{\cal T}_1^b/B}\cos(\phi_a+\varphi_2-\varphi_3)
2\hat{M}^0_2\hat{M}^0_3\,\Big\}\sin\phi_{S_A}\,D_{\pi/g}\nonumber\\
&&+\,\Big\{\,\Big[\,{\rm Im}F_{\!\!a\,+-}^{\,+-}f_{b/B}\cos(\phi_a+\varphi_1-\varphi_2)
\hat{M}^0_1\hat{M}^0_2\nonumber\\
&&\quad\quad +\frac{1}{4}\Delta^{\!N}f_{a/A^\uparrow}\Delta^{\!N}f_{{\cal T}_1^b/B}
\cos(\phi_a-\varphi_1+\varphi_3)\hat{M}^0_1\hat{M}^0_3
\,\Big]\sin(\phi_{S_A}-2\phi_\pi^H)\nonumber\\
&&\quad-\Big[\,{\rm Im}F_{\!\!a\,+-}^{\,-+}
f_{b/B}\cos(\phi_a-\varphi_1+\varphi_2)\hat{M}^0_1\hat{M}^0_2\nonumber\\
&&\quad\quad +\frac{1}{4}\Delta^{\!N}f_{a/A^\uparrow}\Delta^{\!N}f_{{\cal T}_1^b/B}
\cos(\phi_a+\varphi_1-\varphi_3)\hat{M}^0_1\hat{M}^0_3\,\Big]
\sin(\phi_{S_A}+2\phi_\pi^H)\,\Big\}
\Delta^{\!N}D_{\pi/{\cal T}_1^g}\,.
\label{delta-sig-gg}
\end{eqnarray}

The structure is the same as for the quark case, but this time all distributions
related to transversely polarized quark partons and the Collins fragmentation
functions are replaced by analogous functions for linearly polarized gluons,
see Ref.~\cite{Anselmino:2005sh} for more details. Notice that for linearly polarized gluons
inside the polarized hadron $A$, to avoid confusion with notation we prefer to keep
the definitions in terms of the functions
$F_{\lambda^{}_A,\lambda^{\prime}_A}^{\lambda^{}_a,\lambda^{\prime}_a}$.
It is also important to notice that this time the possible
azimuthal asymmetries in the distribution of leading pions inside
the (gluon) jet are proportional to $\cos 2\phi_\pi^H$ and $\sin(\phi_{S_A}\mp2\phi_\pi^H)$,
respectively for the unpolarized and single polarized case.
Therefore, by measuring these asymmetries one
should in principle be able to select contributions coming from either quark or
gluon jet fragmentation.

 It is easy to see that in the case of collinear
initial partons Eq.~(\ref{dsig-unpol-gg}) reduces again to the usual
collinear contribution to the unpolarized cross section, while
Eq.~(\ref{delta-sig-gg}) vanishes. Therefore, the measurement of
such types of asymmetries would be a clear indication that effects
originating from intrinsic parton motion in the initial colliding
hadrons are at work. From the phenomenological point of view,
this would be a crucial test for the TMD approach, independently
of the open issues concerning factorization and universality of the
TMD distribution functions mentioned in the introduction.

Expressions similar to those shown above for the $qq\to qq$
and $gg\to gg$ channels hold also for all partonic contributions
involved, with the appropriate combinations of quark and gluon
distribution and fragmentation functions. In general, less
terms are present both in the denominator and the numerator
of the asymmetry. Moreover, as a general rule quark (gluon) distribution
and fragmentation functions off-diagonal in the parton helicity
indexes, therefore associated to transversely(linearly) polarized
quarks(gluons), appear only in couple. This limits the
number of allowed terms.

According to these results, the single transverse polarized cross section
for the process $A(S_A) B\to {\rm jet}+\pi+X$ will have
the following general structure:
\begin{eqnarray}
2{\rm d}\sigma(\phi_{S_A},\phi_\pi^H) &\sim&  {\rm d}\sigma_0
+{\rm d}\Delta\sigma_0\sin\phi_{S_A}+
{\rm d}\sigma_1\cos\phi_\pi^H+
{\rm d}\Delta\sigma_{1}^{-}\sin(\phi_{S_A}-\phi_\pi^H)+
{\rm d}\Delta\sigma_{1}^{+}\sin(\phi_{S_A}+\phi_\pi^H)\nonumber\\
&+&{\rm d}\sigma_2\cos2\phi_\pi^H
+{\rm d}\Delta\sigma_{2}^{-}\sin(\phi_{S_A}-2\phi_\pi^H)+
{\rm d}\Delta\sigma_{2}^{+}\sin(\phi_{S_A}+2\phi_\pi^H)\,.
\label{d-sig-phi-SA}
\end{eqnarray}

Equivalently, the numerator and denominator of the
asymmetry will have the following expression:
\begin{eqnarray}
&&{\rm d}\sigma(\phi_{S_A},\phi_\pi^H)-
{\rm d}\sigma(\phi_{S_A}+\pi,\phi_\pi^H)\nonumber\\
&&\sim {\rm d}\Delta\sigma_0\sin\phi_{S_A}+
{\rm d}\Delta\sigma_{1}^{-}\sin(\phi_{S_A}-\phi_\pi^H)+
{\rm d}\Delta\sigma_{1}^{+}\sin(\phi_{S_A}+\phi_\pi^H)\nonumber\\
&&+ \;{\rm d}\Delta\sigma_{2}^{-}\sin(\phi_{S_A}-2\phi_\pi^H)+
{\rm d}\Delta\sigma_{2}^{+}\sin(\phi_{S_A}+2\phi_\pi^H)\,,
\label{num-asy-gen}
\end{eqnarray}
\begin{equation}
{\rm d}\sigma(\phi_{S_A},\phi_\pi^H)+
{\rm d}\sigma(\phi_{S_A}+\pi,\phi_\pi^H)
 \equiv 2{\rm d}\sigma^{\rm unp}(\phi_\pi^H) \sim
{\rm d}\sigma_0 + {\rm d}\sigma_1\cos\phi_\pi^H+
{\rm d}\sigma_2\cos2\phi_\pi^H\,.
\label{den-asy-gen}
\end{equation}

In terms of the polarized cross section, Eq.~(\ref{d-sig-phi-SA}), we can define
average values of appropriate circular functions of $\phi_{S_A}$ and $\phi_\pi^H$,
in order to single out the different contributions of interest:
\begin{equation}
\langle\,W(\phi_{S_A},\phi_\pi^H)\,\rangle(\bm{p}_{\rm j},z,k_{\perp\pi})=
\frac{\int{\rm d}\phi_{S_A}{\rm d}\phi_\pi^H\,
W(\phi_{S_A},\phi_\pi^H)\,{\rm d}\sigma(\phi_{S_A},\phi_\pi^H)}
{\int{\rm d}\phi_{S_A}{\rm d}\phi_\pi^H{\rm\, d}\sigma(\phi_{S_A},\phi_\pi^H)}\,.
\label{average}
\end{equation}

Alternatively, for the single spin asymmetry we can, in close analogy with the case of semi-inclusive
deeply inelastic scattering, define appropriate azimuthal moments,
\begin{eqnarray}
A_N^{W(\phi_{S_A},\phi_\pi^H)}(\bm{p}_{\rm j},z,k_{\perp\pi})
&\equiv&
2\langle\,W(\phi_{S_A},\phi_\pi^H)\,\rangle(\bm{p}_{\rm j},z,k_{\perp\pi})\nonumber\\
&=&
2\,\frac{\int{\rm d}\phi_{S_A}{\rm d}\phi_\pi^H\,
W(\phi_{S_A},\phi_\pi^H)\,[{\rm d}\sigma(\phi_{S_A},\phi_\pi^H)-
{\rm d}\sigma(\phi_{S_A}+\pi,\phi_\pi^H)]}
{\int{\rm d}\phi_{S_A}{\rm d}\phi_\pi^H\,
[{\rm d}\sigma(\phi_{S_A},\phi_\pi^H)+
{\rm d}\sigma(\phi_{S_A}+\pi,\phi_\pi^H)]}\,,
\label{gen-mom}
\end{eqnarray}
where $W(\phi_{S_A},\phi_\pi^H)$ is again some appropriate circular function of $\phi_{S_A}$
and $\phi_\pi^H$. In practice, it will be any of the circular functions appearing
e.g.~in Eqs.~(\ref{delta-sig-qq-2}),~(\ref{delta-sig-gg}) for specific partonic channels,
and for polarized cross sections in general
in Eq.~(\ref{num-asy-gen}) so that the coefficient
related to the corresponding azimuthal moment is singled out.
\section{\label{sec-results} Phenomenology}
In this section we will present and discuss some phenomenological implications
of our approach for
the unpolarized and single-transverse polarized case in kinematical
configurations accessible at RHIC by the STAR and PHENIX experiments.
We will consider both central ($\eta_{\rm j}=0$) and forward
($\eta_{\rm j}=3.3$) (pseudo)rapidity configurations
and different c.m.~energies, $\sqrt{s}=62.4$, 200, 500 GeV,
aiming at a check of the potentiality of the approach in
disentangling among different quark and gluon originating effects.
We will also consider two very different situations as far as concern
the TMD distribution and fragmentation functions involved.

We will first consider, for $\pi^+$ production only,
a scenario in which the effects of
all TMD functions are over-maximized. By this we mean that all TMD
functions are maximized in size by imposing natural positivity bounds
(and the Soffer bound for transversity~\cite{Soffer:1994ww,Bacchetta:1999kz});
moreover, the relative signs of
all active partonic contributions are chosen so that they
sum up additively.
This very extreme scenario of course might imply the violation of
other, more stringent, bounds and sum rules;
examples are the Burkardt sume rule for the Sivers
distribution~\cite{Burkardt:2004ur}, and the Sch\"{a}fer-Teryaev
sum rule for the Collins function~\cite{Schafer:1999kn}.
On the other hand, it has the advantage of setting
an upper bound on the absolute value of any of the effects playing
a potential role in the azimuthal asymmetries.
Therefore, all effects that are negligible or even
marginal in this scenario may be directly discarded in subsequent
refined phenomenological analyses.

As a second step in our study we will consider, for
both neutral and charged pions,
only the surviving effects,
involving TMD functions for which parameterizations
are available from independent fits to other spin and azimuthal
asymmetries data in SIDIS, DY, and $e^+e^-$ processes.
Although in our approach factorization and universality are not guaranteed
for the process under consideration,
we still believe that at the present stage this analysis
can be of phenomenological relevance.
It can certainly help in pointing out inconsistencies among
fits based on different processes,
that could be a signal of universality-breaking
effects. On the contrary, good consistency among fits to different observables
from SIDIS, $e^+e^-$ and hadronic collisions data, while not proving
factorization, might signal the smallness of possible universality
breaking terms and the usefulness of the factorization hypothesis in the
present phenomenological analyses.

In this paper, for numerical calculations all TMD distribution and
fragmentation functions will be taken in the simplified form where
the functional dependences on
the parton light-cone momentum fraction and on transverse
motion are completely factorized, assuming a Gaussian-like flavour-independent
shape for the transverse momentum component.
Preliminary lattice QCD calculations seem to support the validity of this
assumption, see e.g.~Ref.~\cite{Hagler:2009ni}.
Notice however that kinematical cuts introduced to prevent,
as usual in the parton model, that the parton longitudinal
momentum (energy) be opposite to (larger than) that of the parent hadron,
effectively lead to a correlation between the light-cone
momentum fraction and the transverse momentum, particularly at
very small and very large ($\to 1$) light-cone momentum fractions
(see e.g.~appendix A of Ref.~\cite{D'Alesio:2004up}).

For the generic parton $a$, the unpolarized and
any of the polarized functions  (that is, the Sivers
and Boer-Mulders distributions and the Collins FF,
and the analogous ones for gluons)
will therefore assume respectively the form
${\cal F}^{\rm unp}_a(u,p)=f^{\rm unp}_a(u)g^{(0)}(p)$ and
$\Delta{\cal F}_a(u,p)=\Delta f_a(u)g^{(i)}(p)$, with $a=q,\bar{q},g$, $i=0,1,2,3$, $u=x$ or $z$ and
$p=k_\perp$ or $k_{\perp\pi}$ for distribution/fragmentation functions
respectively. Our parameterizations  are required to respect angular momentum
conservation in the forward direction, Eqs.~(\ref{kt-power}),~(\ref{kt-power-D});
therefore we define
\begin{equation}
g^{(i)}(p) = \Bigl(\frac{p}{M}\Bigr)^i\,h^{(i)}(p)\,.
\label{gip}
\end{equation}
In particular, $g^{(0)}(p)\equiv g(p)$ is a simple Gaussian normalized to unity:
\begin{equation}
g^{(0)}(p) \equiv g(p) = \frac{1}{\pi\langle\,p^2_0\,\rangle}\,
\exp\bigl[\,-p^2/\langle\,p_0^2\,\rangle\,\bigr]\,,
\label{g0p}
\end{equation}
while
\begin{equation}
g^{(i)}(p) = \left(\frac{p}{M}\right)^i\,h^{(i)}(p) =
K_i\,\left(\frac{p}{M}\right)^i\,\exp\bigl[\,-p^2/\langle\,p_i^2\,\rangle\,\bigr]\,,
\qquad i=1,2,3\,.
\label{hip}
\end{equation}
All the polarized TMD functions are required to fulfill natural positivity bounds
(for transversity, the Soffer bound) with respect to the correponding
unpolarized functions, coming from their general definition as
\begin{equation}
\frac{{\cal F}(S)-{\cal F}(-S)}{{\cal F}(S)+{\cal F}(-S)} =
 \frac{\Delta{\cal F}(S)}{n{\cal F}^{\rm unp}}\,,
\label{delta-F}
\end{equation}
where $S$ is here the spin of the polarized quark or hadron involved, and
$n=1,2$, depending wether the polarized particle is respectively the final/initial one
in the soft process considered (analogous relations
hold for gluons). The positivity bound therefore reads:
\begin{equation}
\frac{|\Delta{\cal F}(u,p)|}{n {\cal F}^{\rm unp}(u,p)} \leq 1\qquad \forall\;  u, p\,.
\label{pos-bound}
\end{equation}
As a matter of fact, to simplify relations we will consider a more conservative
and stringent bound on the two factored components,
\begin{equation}
\frac{|\Delta f_a(u)|}{n f_a^{\rm unp}(u)} \leq 1\qquad \forall\;u,\qquad\qquad
\frac{g^{(i)}(p)}{g^{(0)}(p)} \leq 1 \qquad\forall\; p\,.
\label{bound-2}
\end{equation}
The first condition is usually fulfilled by defining
\begin{equation}
\Delta f_a(u) = n {\cal N}_a(u) f_a^{\rm unp}(u) = n N_a u^{\alpha_a}(1-u)^{\beta_a}
\frac{(\alpha_a+\beta_a)^{(\alpha_a+\beta_a)}}
{\alpha_a^{\alpha_a}\beta_a^{\beta_a}}\,f_a^{\rm unp}(u)\,,
\qquad \mbox{\rm with}\,\,|N_a| \leq 1\,.
\label{f-bound}
\end{equation}
Concerning the transverse momentum dependent component, $g^{(i)}(p)$,
the positivity bound, Eq.~(\ref{bound-2}), can only be
fulfilled if, in Eq.~(\ref{hip}), $\langle\, p_i^2 \,\rangle < \langle\, p_0^2 \,\rangle$.
We then fix the factors $K_i$ in Eq.~(\ref{hip}) by saturating
the bound at the maximum value of $g^{(i)}$. Finally, once a choice has been performed
for $\langle\,p_0^2\,\rangle$, the $\langle\,p_i^2\,\rangle$ are fixed
by maximizing the corresponding $(i+1)$-th $p$-moments.
It is then an easy exercise to verify that this gives the conditions:
\begin{equation}
\langle\,p_1^2\,\rangle = \frac{2}{3}\,\langle\,p_0^2\,\rangle\,,\qquad
\langle\,p_2^2\,\rangle = \frac{1}{2}\,\langle\,p_0^2\,\rangle\,,\qquad
\langle\,p_3^2\,\rangle = \frac{2}{5}\,\langle\,p_0^2\,\rangle\,.
\label{ri}
\end{equation}

{}From the above equations it is clear that the over-maximized
scenario for $\pi^+$ production
we are going to present is obtained by taking, for all polarized
TMD functions, the coefficients ${\cal N}_a(u) = 1$.
For the  transverse momentum component, the procedure delineated above
guarantees the correct power-like behaviour in the forward direction, while
maximizing the moments of the involved functions.

One also needs LO parameterizations for the usual unpolarized
collinear distribution and fragmentation functions. Concerning
the parton distribution functions, we will adopt the unpolarized
set GRV98~\cite{Gluck:1998xa} and (for the Soffer bound) the corresponding longitudinally
polarized set GRSV2000~\cite{Gluck:2000dy}. Since the range of the jet transverse momentum
(the hard scale) covered is significant, we will take into account proper evolution
with scale. Concerning transversity, in the maximized scenario we will fix
it at the initial scale by saturating the Soffer bound and then letting it
evolve. On the other hand, the transverse momentum component of all TMD functions
is kept fixed with no evolution with scale.
Notice that at this stage evolution properties of the full TMD functions are not
known.

As for fragmentation functions, we will adopt two well-known LO sets
among those available in the literature, the set by Kretzer
(K)~\cite{Kretzer:2000yf} and the one by
De Florian, Sassot and Stratmann (DSS)~\cite{deFlorian:2007aj}.
Our choice is dictated by the subsequent
use of the two available parametrization sets for the Sivers and Collins functions
in our scheme, that have been derived in the past years by adopting these sets of FFs.
Let us notice that, as the authors of Ref.~\cite{deFlorian:2007aj} suggest,
 the LO DSS fragmentation function set has to be handled with some care.
In fact, aiming at reproducing for the first time unpolarized cross sections for inclusive
hadronic collisions, at LO accuracy a very huge gluon component (as compared to other
FF sets) is required. This could be an artifact of the LO set,
which in fact is sizably reduced in the NLO parameterizations.
However, we are at present forced to work at leading order in our TMD approach.
The comparison among the two sets adopted will allow to stress
the possible effects of the large gluon component in the LO DSS set.

Concerning the parameterizations of the transversity and Sivers distributions, and
of the Collins functions, we will consider two sets resulting from the fits to available
data on azimuthal asymmetries in polarized SIDIS from HERMES and COMPASS experiments,
and on hadron-pair production in $e^+e^-$ collisions from Belle:\\
Set 1 (SIDIS~1) includes the $u$, $d$ quark Sivers functions of Ref.~\cite{Anselmino:2005ea},
the $u$, $d$ quark transversity distributions and the favoured
and unfavoured Collins FFs of Ref.~\cite{Anselmino:2007fs}.
Data include preliminary HERMES data  for charged pions~\cite{Diefenthaler:2005gx}
and COMPASS data on charged hadrons with a deuteron target~\cite{Alexakhin:2005iw}
on the SIDIS Sivers asymmetry;
HERMES data for charged pions~\cite{Airapetian:2004tw,Pappalardo:2008zz}
and COMPASS data for charged hadrons with a deuteron target~\cite{Ageev:2006da}
on the SIDIS Collins asymmetry; early Belle data on azimuthal asymmetries
for hadron-pair production in $e^+e^-$ collisions~\cite{Abe:2005zx};
no SIDIS data on kaons were used and the Kretzer
set~\cite{Kretzer:2000yf} for pion FFs was used.\\
Set 2 (SIDIS~2) includes the new $u$, $d$, and sea-quark
Sivers functions of Ref.~\cite{Anselmino:2008sga} and an updated set
of the $u$, $d$ quark transversity distributions and of the favoured
and unfavoured Collins FFs of Ref.~\cite{Anselmino:2008jk}.
Corresponding data include:
preliminary HERMES data on pions and charged kaons~\cite{Diefenthaler:2007rj}
and preliminary COMPASS data on charged pions and kaons
with a deuteron target~\cite{Martin:2007au} for the SIDIS Sivers asymmetry;
preliminary HERMES data for pions~\cite{Diefenthaler:2007rj}
and COMPASS data for charged pions with a deuteron target~\cite{Alekseev:2008dn}
on the SIDIS Collins asymmetry; recent Belle data on azimuthal asymmetries
for hadron-pair production in $e^+e^-$ collisions~\cite{Seidl:2008xc};
the DSS set~\cite{deFlorian:2007aj} for pion and kaon FFs was adopted.

Notice that the almost unknown gluon Sivers function was tentatively
taken positive and saturated to an updated version of the bound obtained
in Ref.~\cite{Anselmino:2006yq} by considering
PHENIX data for the $\pi^0$ transverse SSA at mid-rapidity production in
polarized $pp$ collisions at RHIC~\cite{Adler:2005in}.

It is also important to stress here that polarized SIDIS data on azimuthal
asymmetries from HERMES and COMPASS experiments cover a relatively limited
range of Bjorken $x$, $x_{\rm B}\leq 0.3$. Therefore, the statistical uncertainty
of the parameterizations available for the transversity and Sivers
distributions is huge  at large $x$ values, where one extrapolates their behaviour.
As we will see, this reflects in the very different behaviour of the Sivers
and Collins asymmetries when estimated adopting sets SIDIS~1 and SIDIS~2 respectively.

Finally, regarding the quark Boer-Mulders distribution function much less is known and
available parameterizations have large uncertainties. In our calculations
we have adopted the recent parametrization by
Barone, Melis and Prokudin (BMP)~\cite{Barone:2009hw}, which makes use of
our Set SIDIS~2 for the transversity and Sivers distributions and for the
Collins function.

We have considered the following kinematical configurations for the
PHENIX and STAR experiments at RHIC:\\
1) $\sqrt{s}=62.4$ GeV, $\eta_{\rm j}=0$, $1\leq p_{{\rm j}\,T} \leq 14$ GeV;\\
2) At $\sqrt{s}=62.4$ GeV, the forward rapidity configuration covers a very limited
range of $p_{{\rm j}\,T}$ values which probably prevents the unambiguous definition of jets and
leading particles, therefore we will not consider it in the sequel;\\
3) $\sqrt{s}=200$ GeV, $\eta_{\rm j}=0$, $2\leq p_{{\rm j}\,T} \leq 15$ GeV;\\
4) $\sqrt{s}=200$ GeV, $\eta_{\rm j}=3.3$, $2\leq p_{{\rm j}\,T} \leq 6.5$ GeV
 $[0.27\leq x_F\leq 0.88]$;\\
5) $\sqrt{s}=500$ GeV, $\eta_{\rm j}=0$, $2\leq p_{{\rm j}\,T} \leq 15$ GeV;\\
6) $\sqrt{s}=500$ GeV, $\eta_{\rm j}=3.3$, $2\leq p_{{\rm j}\,T} \leq 15$ GeV
 $[0.11\leq x_F\leq 0.81]$.

For completeness, in the forward rapidity case we have also shown the range of
$x_F$ covered, where $x_F$ is the usual Feynman variable for the jet,
$x_F=2 p_{{\rm j}\,L}/\sqrt{s}$.

In all cases considered, since we are interested to azimuthal asymmetries for
leading particles inside the jet, we will present results obtained
integrating the light-cone momentum fraction of the observed hadron, $z$,
in the range $z\geq 0.3$.
\subsection{\label{sec-results-unp} Azimuthal asymmetries for the unpolarized cross section}
In this section we will discuss results for the azimuthal $\langle\cos\phi_\pi^H\rangle$,
$\langle\cos2\phi_\pi^H\rangle$ asymmetries [see Eq.~(\ref{average})]
in the unpolarized cross section for the process
$p p\to {\rm jet}+\pi+X$.

\begin{figure*}[t]
 \includegraphics[angle=0,width=0.4\textwidth]{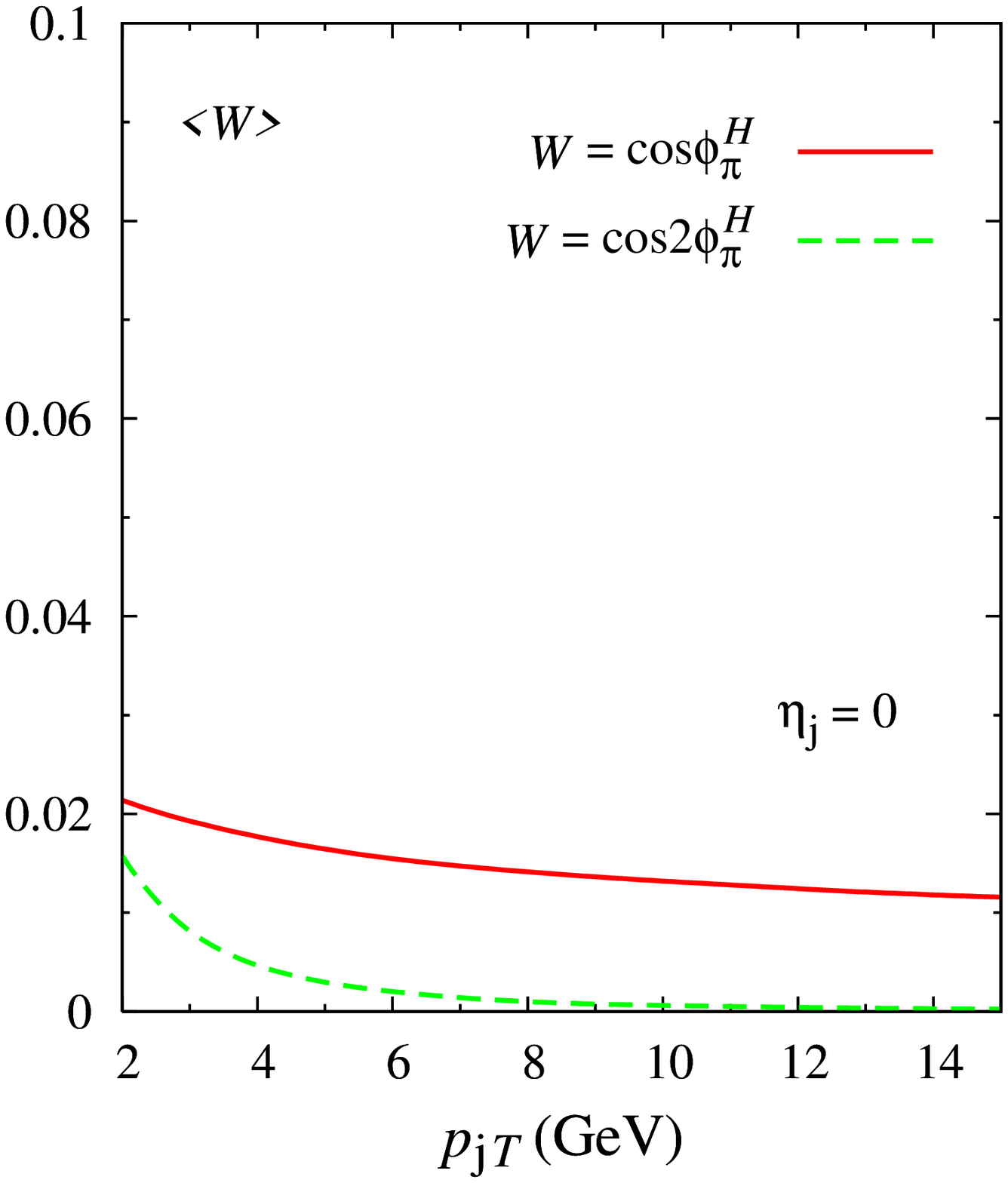}
 \includegraphics[angle=0,width=0.4\textwidth]{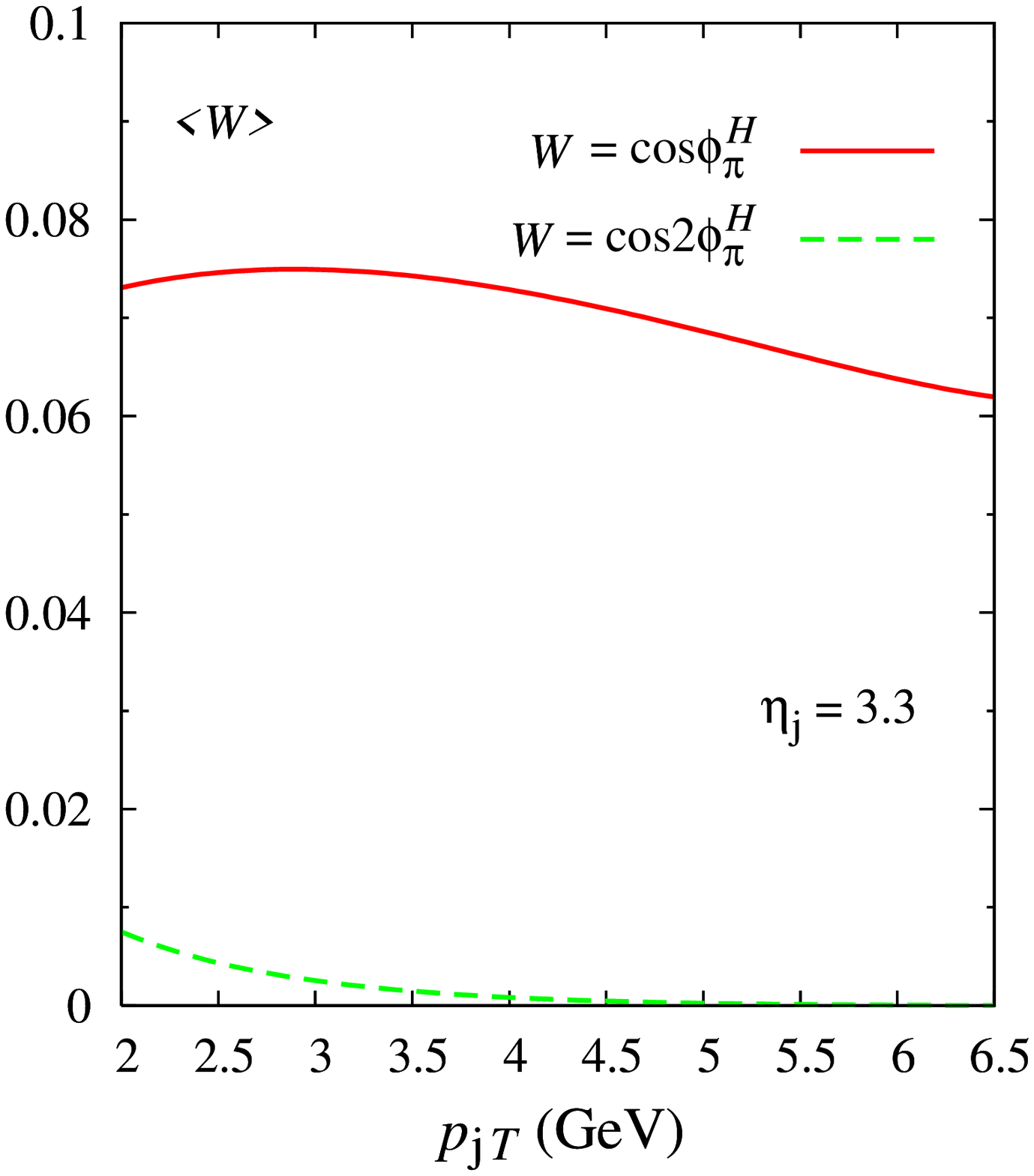}
 \caption{(color online). Maximized  quark-originated ($\cos\phi_\pi^H$) and
 gluon-originated ($\cos2\phi_\pi^H$) asymmetries
(solid-red and dashed-green lines respectively) for the
unpolarized $pp\to {\rm jet}+\pi^+ + X$ process,
at $\sqrt{s}=200$ GeV c.m.~energy in the central (left panel)
and forward (right panel) rapidity region as a function of $p_{{\rm j}\,T}$,
from $p_{{\rm j}\,T}=2$ GeV up to the maximum allowed value, adopting
the Kretzer FF set. Slightly lower(similar) values are obtained
for quark(gluon) asymmetries when using the DSS set.
 \label{asy-unp-200} }
\end{figure*}

As it is easy to verify by looking e.g.~at Eqs.~(\ref{dsig-unpol-qq}),
(\ref{dsig-unpol-gg}) for the $qq\to qq$ and $gg\to gg$ partonic contributions
(analogous results, where allowed, hold for all other channels),
in the unpolarized case:\\
1) The symmetric part gets contributions by the usual unpolarized term,
already present in the collinear approach, and by an additional term
involving a Boer-Mulders$\,\otimes\,$Boer-Mulders convolution for the initial
quarks (or the analogous terms involving linearly polarized gluons);
however, we have explicitely checked that even  in the maximized
scenario this last contribution is always negligible in all the kinematical
configurations considered;
therefore, we will not discuss it anymore in the sequel;\\
2) The $\cos\phi_\pi^H$ asymmetry is generated by the
quark Boer-Mulders$\,\otimes\,$Collins  convolution term,
involving a transversely polarized quark and an unpolarized hadron
both in the initial state and in the fragmentation process.
In the central rapidity region ($\eta_{\rm j}=0$) the maximized
value of this asymmetry is of the order 1-3\%, depending
on the fragmentation function set adopted and on the c.m.~energy
considered, being almost negligible at $\sqrt{s}=500$ GeV.
In the forward rapidity region, $\eta_{\rm j}=3.3$, the maximized
$\cos\phi_\pi^H$ asymmetry can be much larger both at $\sqrt{s}=200$
and 500 GeV.
As an example, in Fig.~\ref{asy-unp-200} we show the
maximized  $\cos\phi_\pi^H$ asymmetry (solid red lines)
for $\pi^+$ production
at c.m.~energy $\sqrt{s}=200$ GeV  in the central (left panel)
and forward (right panel) rapidity region as a function of $p_{{\rm j}\,T}$,
from $p_{{\rm j}\,T}=2$ GeV up to the maximum allowed value, adopting
the Kretzer FF set. Slightly lower values are obtained using the
DSS set.\\
3) The $\cos2\phi_\pi^H$ asymmetry is related to the term involving
linearly polarized gluons and unpolarized hadrons both in the
initial state and in the fragmentation process, that is the
convolution of a Boer-Mulders-like gluon distribution with
a Collins-like gluon FF. Even the maximized contribution is
practically negligible in the kinematical configurations considered.
As an example, again in Fig.~\ref{asy-unp-200}, we show the
maximized  $\cos2\phi_\pi^H$ asymmetry (dashed green lines)
for $\pi^+$ production
at $\sqrt{s}=200$ GeV c.m.~energy in the central (left panel)
and forward (right panel) rapidity region as a function of $p_{{\rm j}\,T}$,
adopting the Kretzer FF set. Similar results are obtained using the
DSS set.

Concerning results with available parameterizations,
for the quark-originated $\cos\phi_\pi^H$ asymmetry we have verified
that the asymmetries obtained with the parameterizations adopted here, our set
SIDIS~2 and the BMP set for the Boer-Mulders function, are negligible in all
kinematical configurations considered. No parameterizations are presently available
for the analogous gluon contributions leading to the  $\cos2\phi_\pi^H$ asymmetry.
\subsection{\label{sec-results-an} Azimuthal asymmetries for
$A_N(p^\uparrow p\to{\rm jet}+\pi+X)$}
Let us now discuss our numerical results for the Sivers ($A_N^{\sin\phi_{S_A}}$) asymmetry
and the quark [$A_N^{\sin(\phi_{S_A}\mp\phi_\pi^H)}$] and gluon
[$A_N^{\sin(\phi_{S_A}\mp2\phi_\pi^H)}$] Collins(-like) asymmetries,
see Eq.~(\ref{gen-mom}).
Our estimates are qualitatively similar at the three different c.m.~energies
considered, with some differences in the size of the asymmetries
and in the relative weight of the quark and gluon contributions where
both play a role. Therefore, we will concentrate on the results
obtained at $\sqrt{s}=200$ GeV.
\subsubsection{\label{sec-results-an-siv} The Sivers asymmetry}
In this case, both quark and gluon contributions can be present,
and they cannot be disentangled. However, some kinematical configurations
can be dominated by quark or gluons terms and a sizable asymmetry
in these regions might be an unambiguous indication for a Sivers asymmetry
generated by the dominant partonic contribution.

\begin{figure*}[t]
 \includegraphics[angle=0,width=0.4\textwidth]{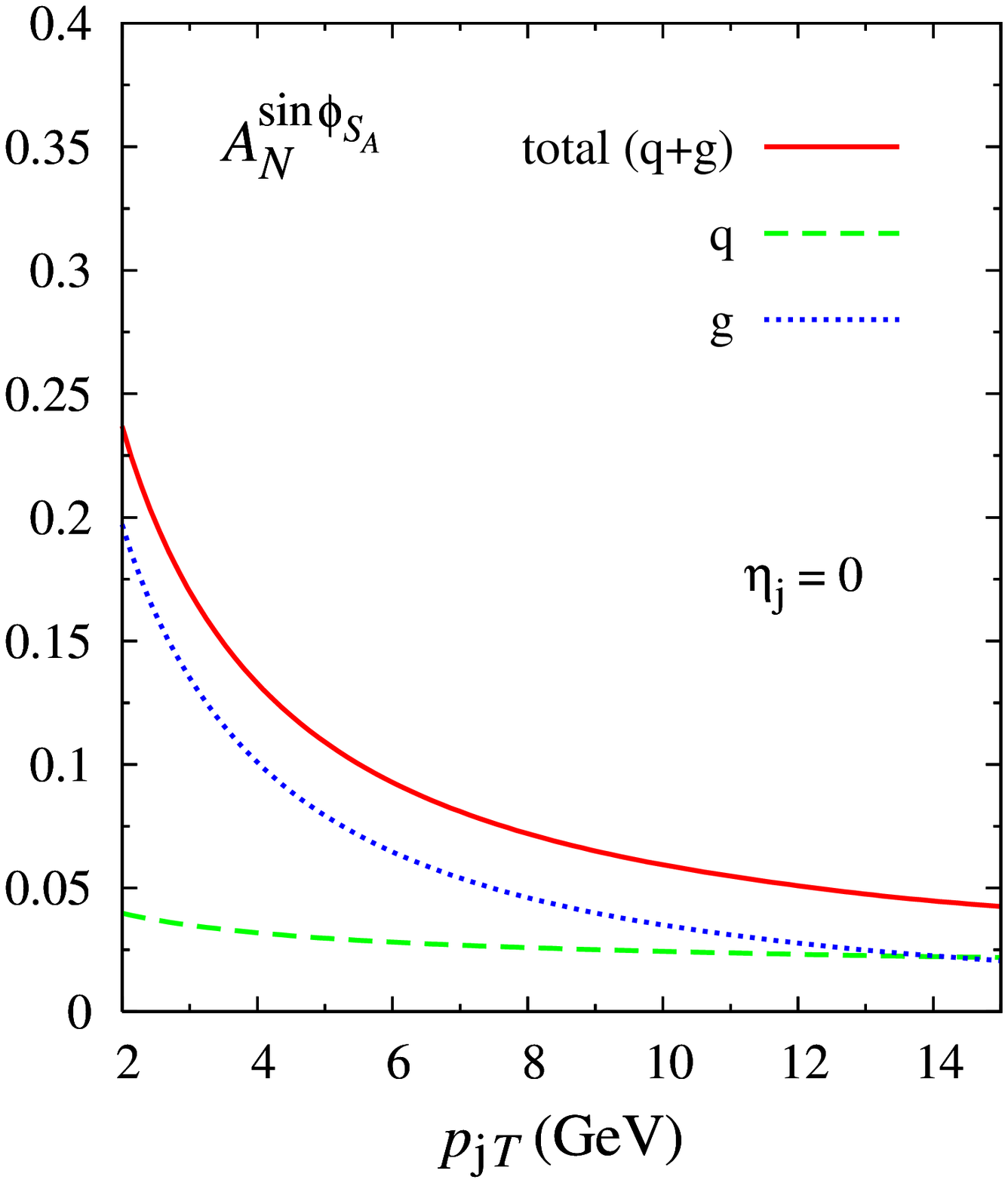}
 \includegraphics[angle=0,width=0.4\textwidth]{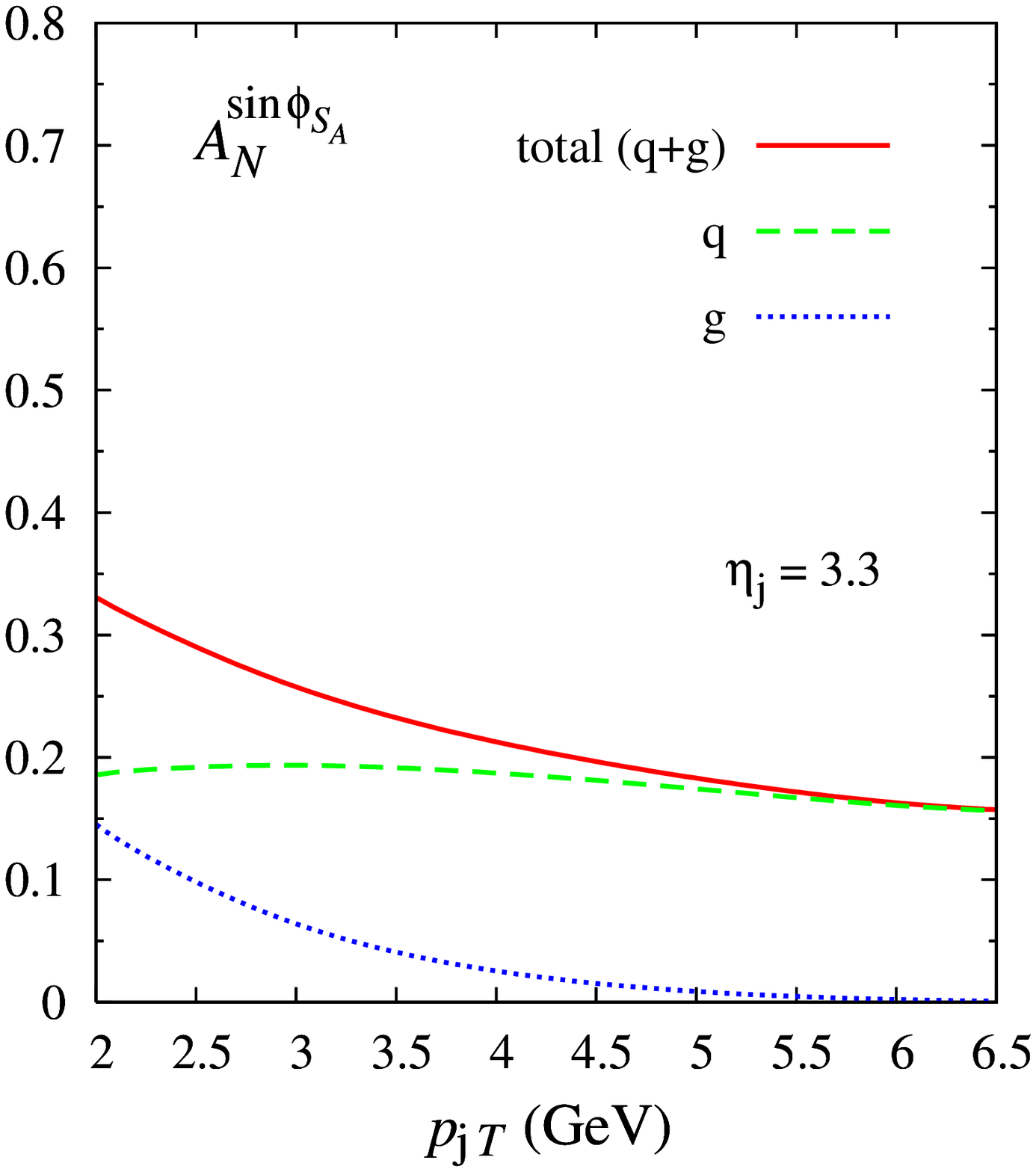}
 \caption{(color online). Maximized  total (solid red line),
quark-originated (dashed green line) and
gluon-originated (dotted blue line) Sivers asymmetry
for the $p^\uparrow p\to {\rm jet}+\pi^+ + X$ process,
at $\sqrt{s}=200$ GeV c.m.~energy in the central (left panel)
and forward (right panel) rapidity region as a function of $p_{{\rm j}\,T}$,
from $p_{{\rm j}\,T}=2$ GeV up to the maximum allowed value, adopting
the Kretzer FF set.
Similar results, with some differences
in the total size and in the relative weight of the quark and
gluon contributions are obtained adopting the DSS set of
fragmentation functions and considering different c.m.~energies.
 \label{asy-an-siv-max200} }
\end{figure*}

In Fig.~\ref{asy-an-siv-max200} we show the total observable Sivers
asymmetry (solid red line), and the corresponding quark and gluon contributions
(dashed green and dotted blue lines respectively) for $\pi^+$ production,
in the maximized scenario and adopting the Kretzer fragmentation function
set, at $\sqrt{s}=200$ GeV and as a function of $p_{{\rm j}\,T}$ in the
central (left panel) and forward (right panel) rapidity regions.
The maximized potential Sivers asymmetry can be very large in both
cases. In the central rapidity region, the asymmetry is dominated by
the gluon contribution at the lowest $p_{{\rm j}\,T}$ range while
gets comparable quark and gluon contributions in the large
$p_{{\rm j}\,T}$ range. A large Sivers asymmetry around
$p_{{\rm j}\,T}=4\div6$ GeV could then be a clear indication for
a sizable gluon contribution. However, one must not forget
that, as mentioned above, recent PHENIX results for $A_N(p^\uparrow p\to\pi^0+X)$
in the central rapidity region put much more stringent bounds
on the gluon Sivers distribution than the simple positivity
bound adopted in the maximized scenario. We have checked
that even adopting this more stringent bound, a potential gluon-generated
Sivers asymmetry of the order of 2\% might survive in this region,
being possibly measurable.
In the forward rapidity region, on the contrary, the quark and
gluon contributions are comparable at low $p_{{\rm j}\,T}$ values,
while the maximized asymmetry is dominated by the quark contribution
for $p_{{\rm j}\,T}\gtrsim 4$ GeV. Therefore, a large Sivers
asymmetry in this kinematical range could be ascribed unambiguosly
to the quark Sivers effect.
Qualitatively similar results are obtained at the other c.m.~energies considered
or adopting the DSS set of fragmentation functions,
with some changes in the total size and in the relative weights of the
quark and gluon contributions.

\begin{figure*}[b]
 \includegraphics[angle=0,width=0.34\textwidth]{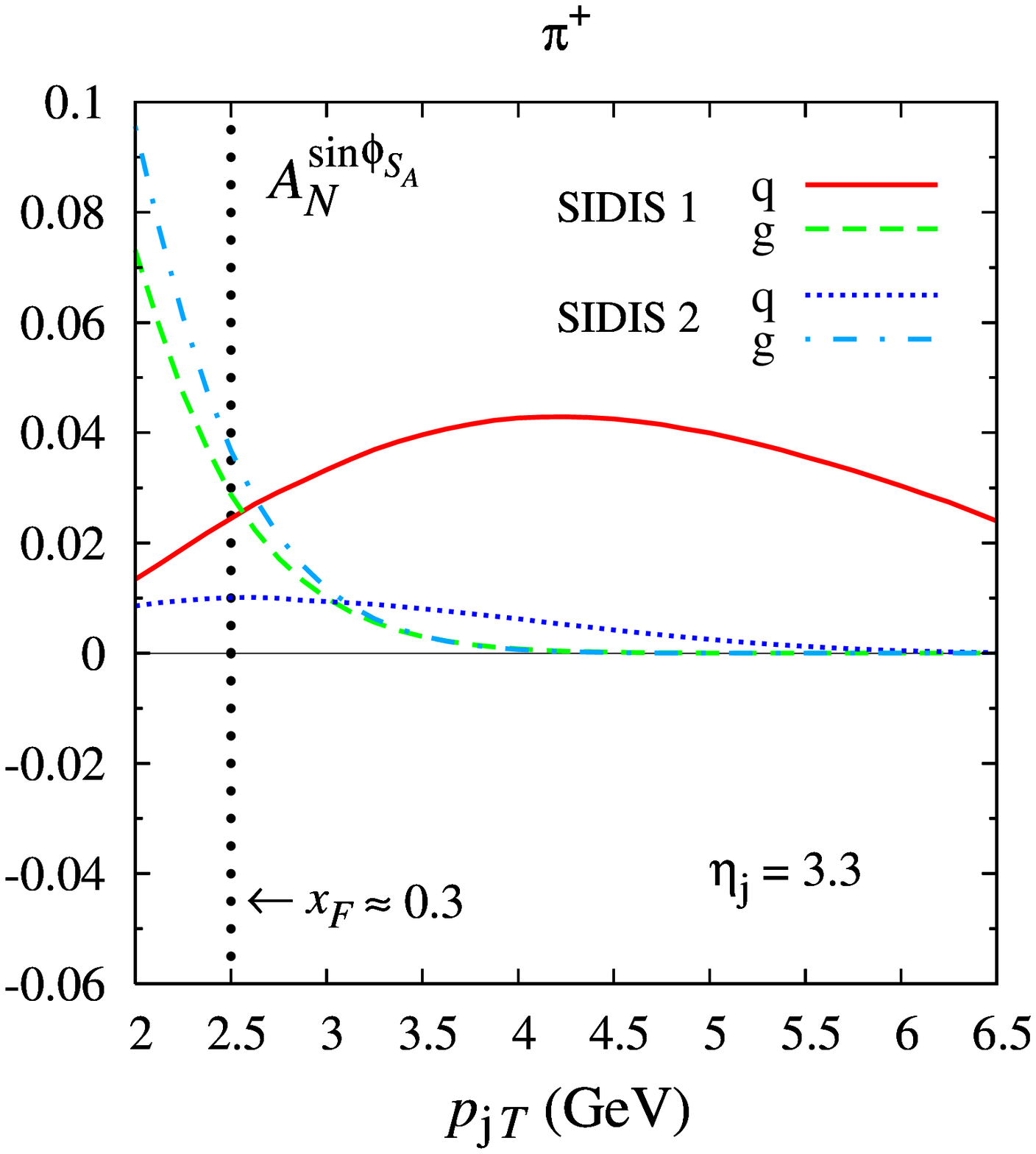}
 \hspace*{-20pt}
 \includegraphics[angle=0,width=0.34\textwidth]{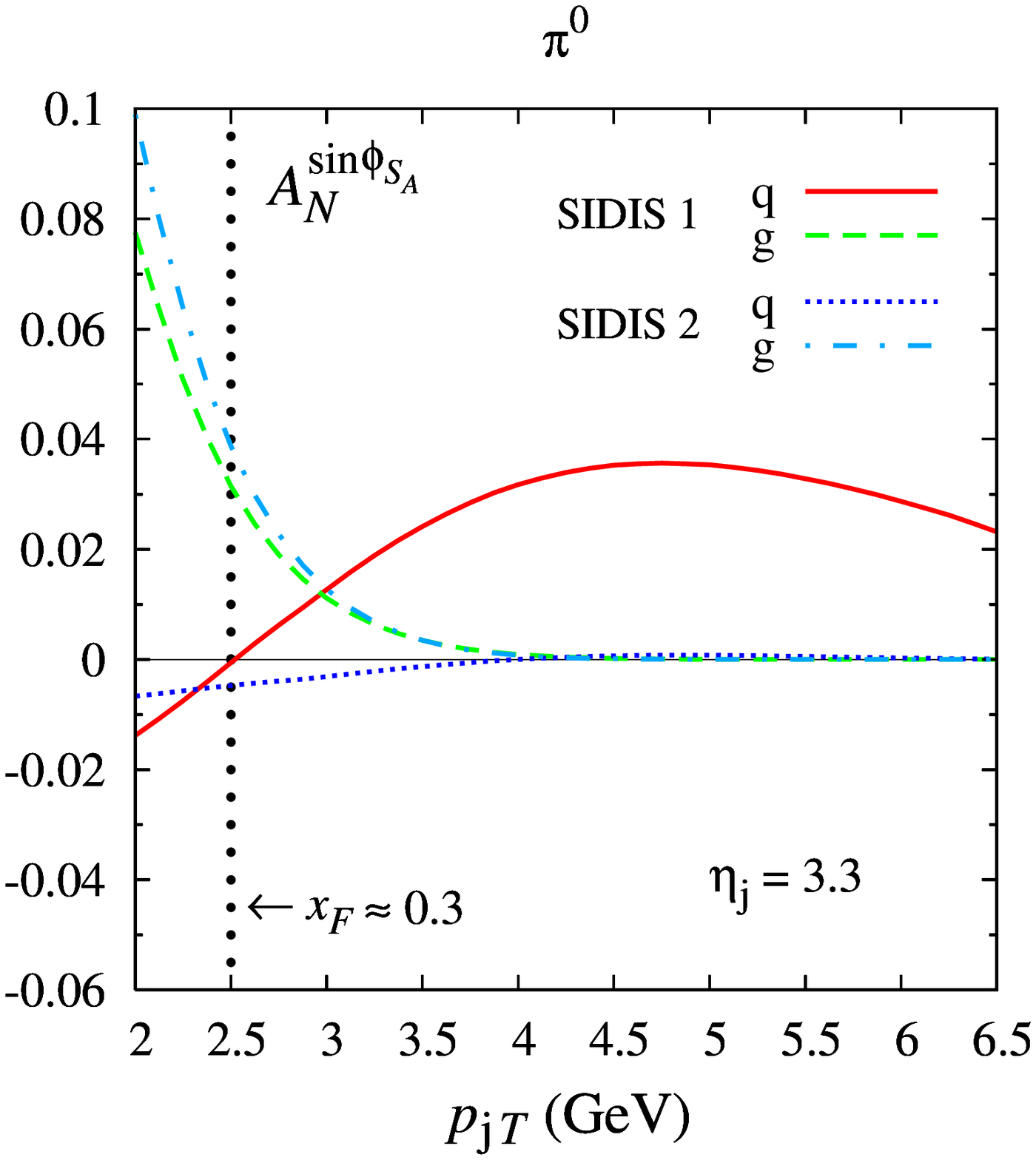}
 \hspace*{-20pt}
 \includegraphics[angle=0,width=0.34\textwidth]{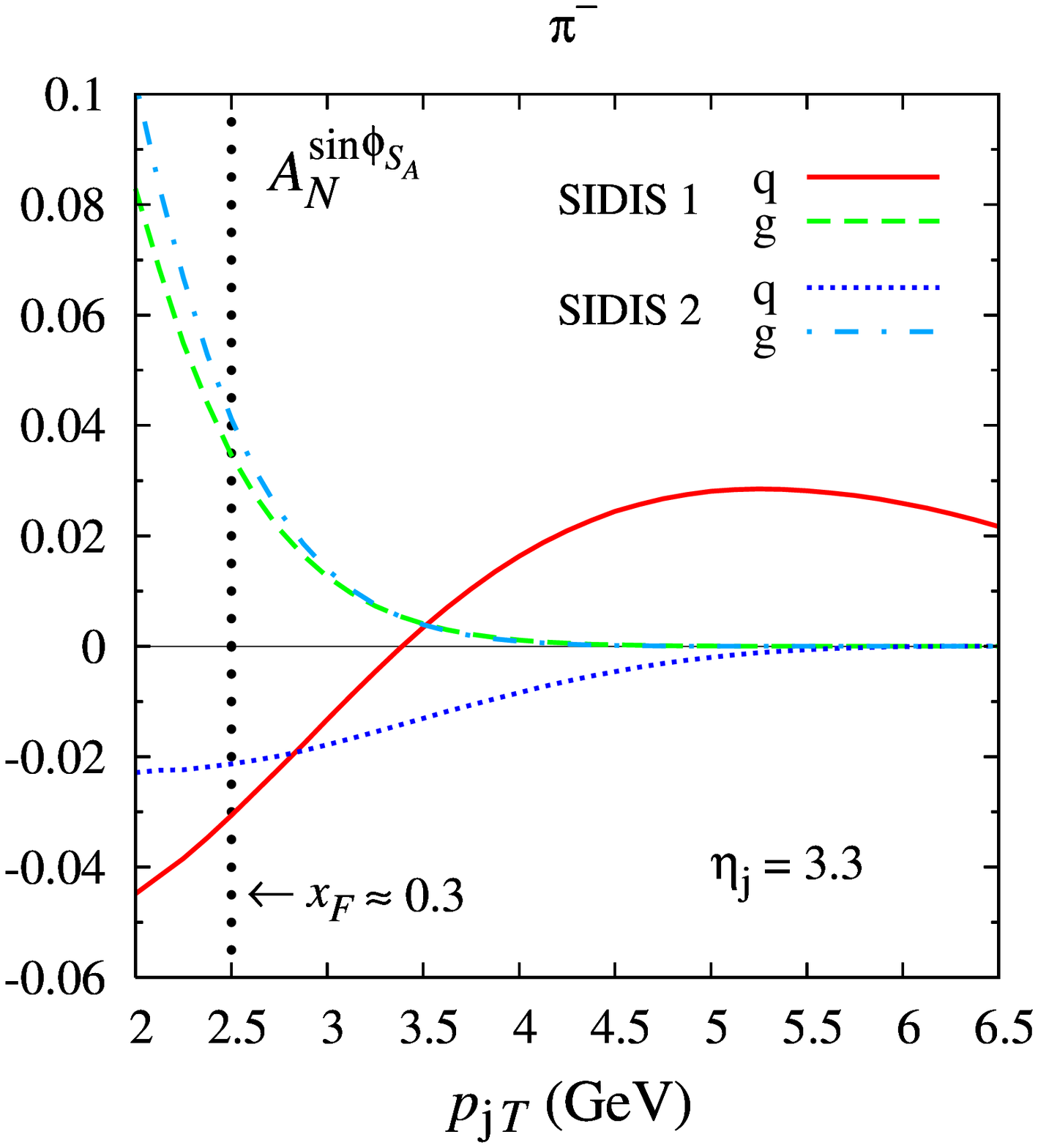}
 \caption{(color online). The estimated quark and gluon
 contributions to the Sivers asymmetry
 for the $p^\uparrow p\to {\rm jet}+\pi + X$ process,
 obtained adopting respectively the parametrization sets
 SIDIS~1 (quark contribution: solid red line;
  gluon contribution: dashed green line) and
 SIDIS~2 (quark contribution: dotted blue line;
gluon contribution: dot-dashed cyan line),
at $\sqrt{s}=200$ GeV c.m.~energy in the forward rapidity
region and as a function of $p_{{\rm j}\,T}$,
from $p_{{\rm j}\,T}=2$ GeV up to the maximum allowed value.
The dotted black vertical line delimits the region
beyond which the SIDIS parameterizations for the quark
Sivers function are presently plagued by large uncertainties.
Similar results are obtained when considering
different c.m.~energies.
 \label{asy-an-siv-par200} }
\end{figure*}

In Fig.~\ref{asy-an-siv-par200} we show,
for both neutral and charged pions, the quark and gluon
contributions to the Sivers asymmetry,
obtained adopting respectively the parametrization sets SIDIS~1
(quark contribution: solid red line; gluon contribution: dashed green line)
and SIDIS~2 (quark contribution: dotted blue line;
gluon contribution: dot-dashed cyan line), and the updated version
of the bound on the gluon Sivers asymmetry derived in Ref.~\cite{Anselmino:2006yq},
at $\sqrt{s}=200$ GeV and in the forward rapidity region, as a function of
$p_{{\rm j}\,T}$. The dotted black vertical line delimits the region
beyond which the SIDIS parameterizations for the quark Sivers
distribution are extrapolated outside the $x_{\rm B}$
region covered by SIDIS data and are therefore plagued by large uncertainties.
This reflects on the fact that below this limit the two sets give comparable
results, while above it they differ remarkably. In particular,
at the largest reachable $p_{{\rm j}\,T}$ values the SIDIS~1 set gives
a Sivers asymmetry of the order $2\div4\%$,  while the SIDIS~2 set
leads to a negligible asymmetry.
Therefore, a measurement of this asymmetry might help in clarifying
the behaviour of the quark Sivers distribution in the large $x$ region,
which is not covered by present SIDIS data from HERMES and COMPASS
experiments. Future planned measurements at the Jefferson Lab (Jlab)
12 GeV Upgrade will also be very useful in this respect
(see e.g.~Ref.~\cite{Gao:2010av}).
\subsubsection{\label{sec-results-an-col}
The Collins $A_N^{\sin(\phi_{S_A}\mp\phi_\pi^H)}$
and the Collins-like $A_N^{\sin(\phi_{S_A}\mp2\phi_\pi^H)}$ asymmetries}
Let us first briefly discuss the quark generated asymmetry
$A_N^{\sin(\phi_{S_A}+\phi_\pi^H)}$. It comes from two distinct
contributions, see e.g.~Eq.~(\ref{delta-sig-qq-2})
for the $qq\to qq$ channel:
one involving the convolution between the term of
the TMD transversity distribution suppressed in the collinear
configuration ($\propto k^2_{\perp q}h_{1T}^{\perp q}$)
and the Collins function; another term
involving the convolution of the Sivers and Boer-Mulders
distributions for the initial quarks with the Collins function
for the final quark [an analogous term appears also
in the $A_N^{\sin(\phi_{S_A}-\phi_\pi^H)}$
asymmetry, see Eq.~(\ref{delta-sig-qq-2})].
We have explicitly checked that for the process under study
and the kinematical configurations considered both these
contributions are always negligible already in the maximized scenario.
Therefore we will not consider the  $\sin(\phi_{S_A}+\phi_\pi^H)$
asymmetry in the sequel. A similar situation holds also for the
gluon generated $A_N^{\sin(\phi_{S_A}+2\phi_\pi^H)}$ asymmetry, where two
contributions analogous to the quark ones discussed above but for
linearly polarized gluons are involved.

\begin{figure*}[t]
 \includegraphics[angle=0,width=0.4\textwidth]{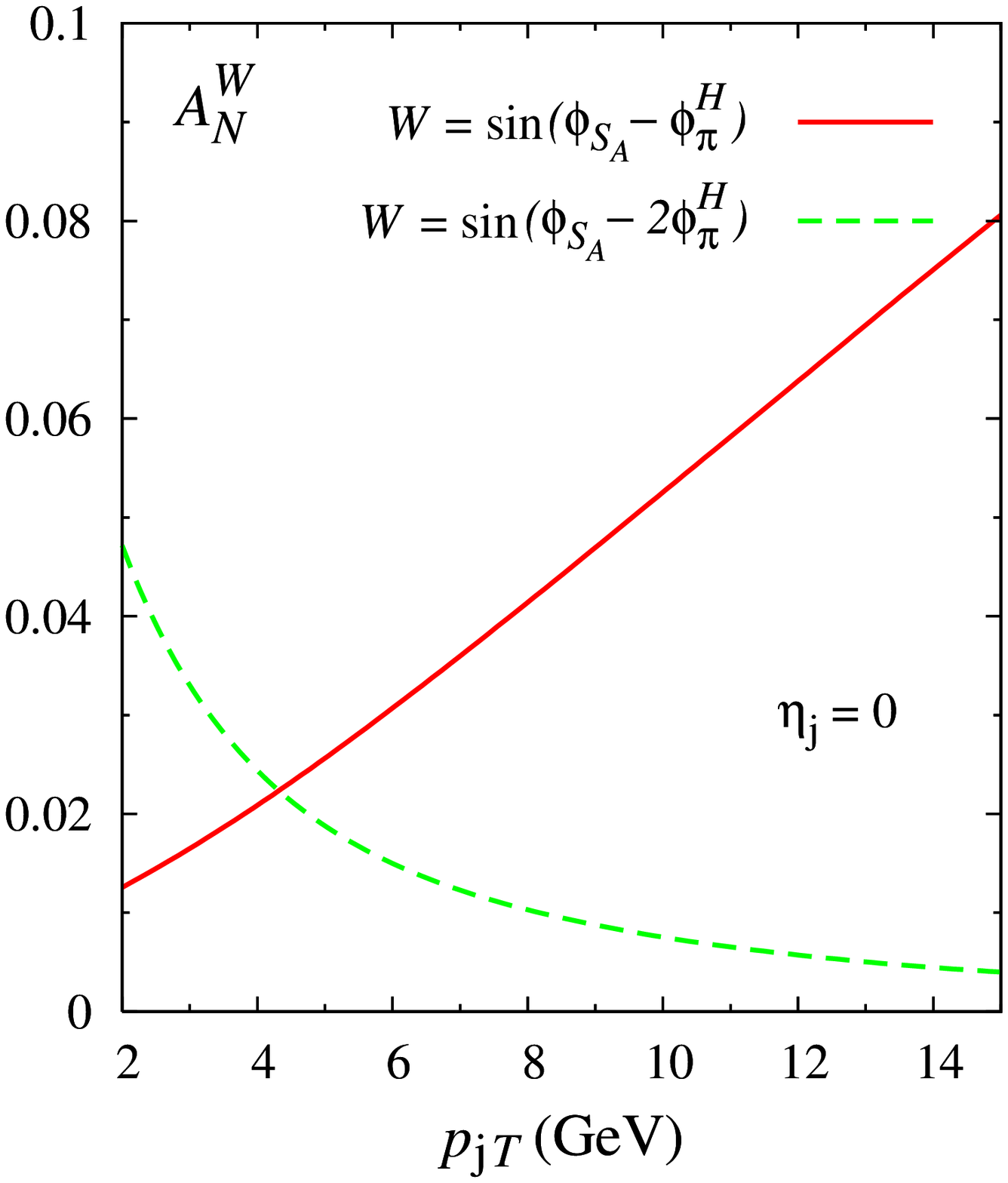}
 \includegraphics[angle=0,width=0.4\textwidth]{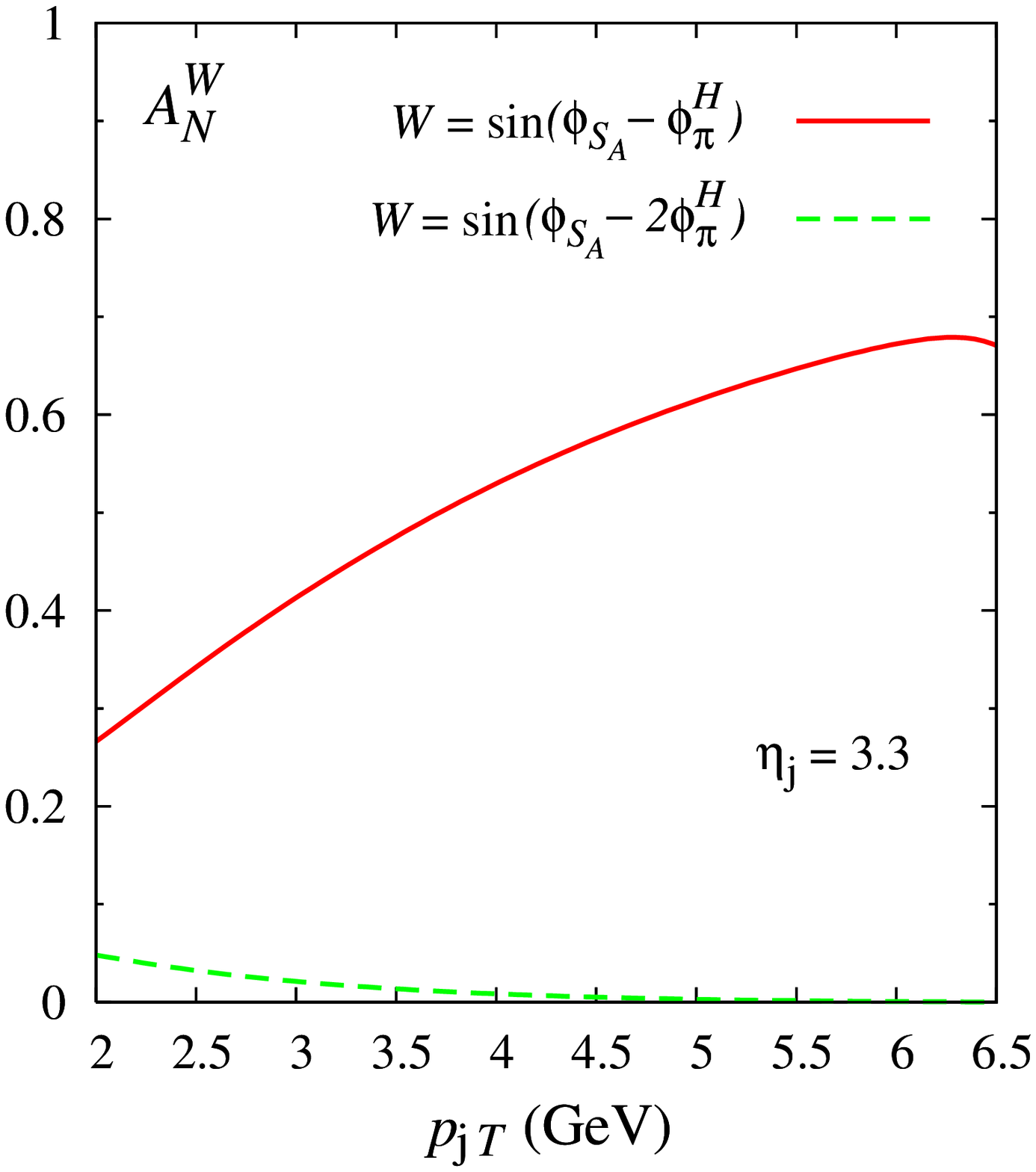}
 \caption{(color online). Maximized  quark (solid red line)
and gluon (dashed green line) Collins(-like) asymmetries
for the $p^\uparrow p\to {\rm jet}+\pi^+ + X$ process,
at $\sqrt{s}=200$ GeV c.m.~energy in the central (left panel)
and forward (right panel) rapidity region as a function of $p_{{\rm j}\,T}$,
from $p_{{\rm j}\,T}=2$ GeV up to the maximum allowed value, adopting
the Kretzer FF set. Notice the difference in the scale
between the two panels.
Similar results, with some differences
in the total size and in the relative weight of the quark and
gluon contributions are obtained adopting the DSS set of
fragmentation functions and considering different c.m.~energies.
 \label{asy-an-coll-max200} }
\end{figure*}

In Fig.~\ref{asy-an-coll-max200}
we present the quark $A_N^{\sin(\phi_{S_A}-\phi_\pi^H)}$ Collins asymmetry
(solid red lines) and the gluon $A_N^{\sin(\phi_{S_A}-2\phi_\pi^H)}$
Collins-like asymmetry (dashed green lines) in the maximized scenario
for the $p^\uparrow p\to {\rm jet}+\pi^+ + X$ process,
at $\sqrt{s}=200$ GeV c.m.~energy in the central (left panel)
and forward (right panel) rapidity region as a function of $p_{{\rm j}\,T}$,
from $p_{{\rm j}\,T}=2$ GeV up to the maximum allowed value, adopting
the Kretzer FF set.
In the central rapidity region
the quark Collins asymmetry is very small at the lowest $p_{{\rm j}\,T}$
values, then increases almost linearly reaching about 8\% at the upper range.
At $\sqrt{s}=500 (62.4)$ GeV c.m.~energy the behaviour is similar
and the largest reached size, at large $p_{{\rm j}\,T}$ values,
is about half (twice) respectively.
Results with the DSS fragmentation function set are slightly lower in size.
Instead, in the forward rapidity region the asymmetry is (potentially)
always large and increases almost linearly from about 25\% to about 70\%
going from the lowest to the largest $p_{{\rm j}\,T}$ values.
Results are very similar at $\sqrt{s}=500$ GeV and when adopting
the DSS fragmentation function set.

Concerning the gluon Collins-like asymmetry, both in the central
and in the forward rapidity regions it is of the order of 5\%
at the lowest $p_{{\rm j}\,T}$ values, then starts decreasing slowly
and becomes negligible at large $p_{{\rm j}\,T}$ values.
Very similar results hold at different energies and when adopting
the DSS set.

Let us now consider, for both neutral and charged pions,
numerical results for the quark Collins
asymmetry obtained adopting the parameterizations SIDIS~1 and
SIDIS~2 for the transversity distribution and the Collins
fragmentation function (no parameterizations are available yet
in the analogous gluon case).
It turns out that in the central rapidity region the estimated
asymmetry is practically negligible in all cases considered
(different c.m.~energies and FF sets). Only for the SIDIS~2
parametrization and at $\sqrt{s}=62.4$ GeV the asymmetry
for charged pions can reach about $2\div3$\% in size
at large $p_{{\rm j}\,T}$ values.

\begin{figure*}[t]
 \includegraphics[angle=0,width=0.4\textwidth]{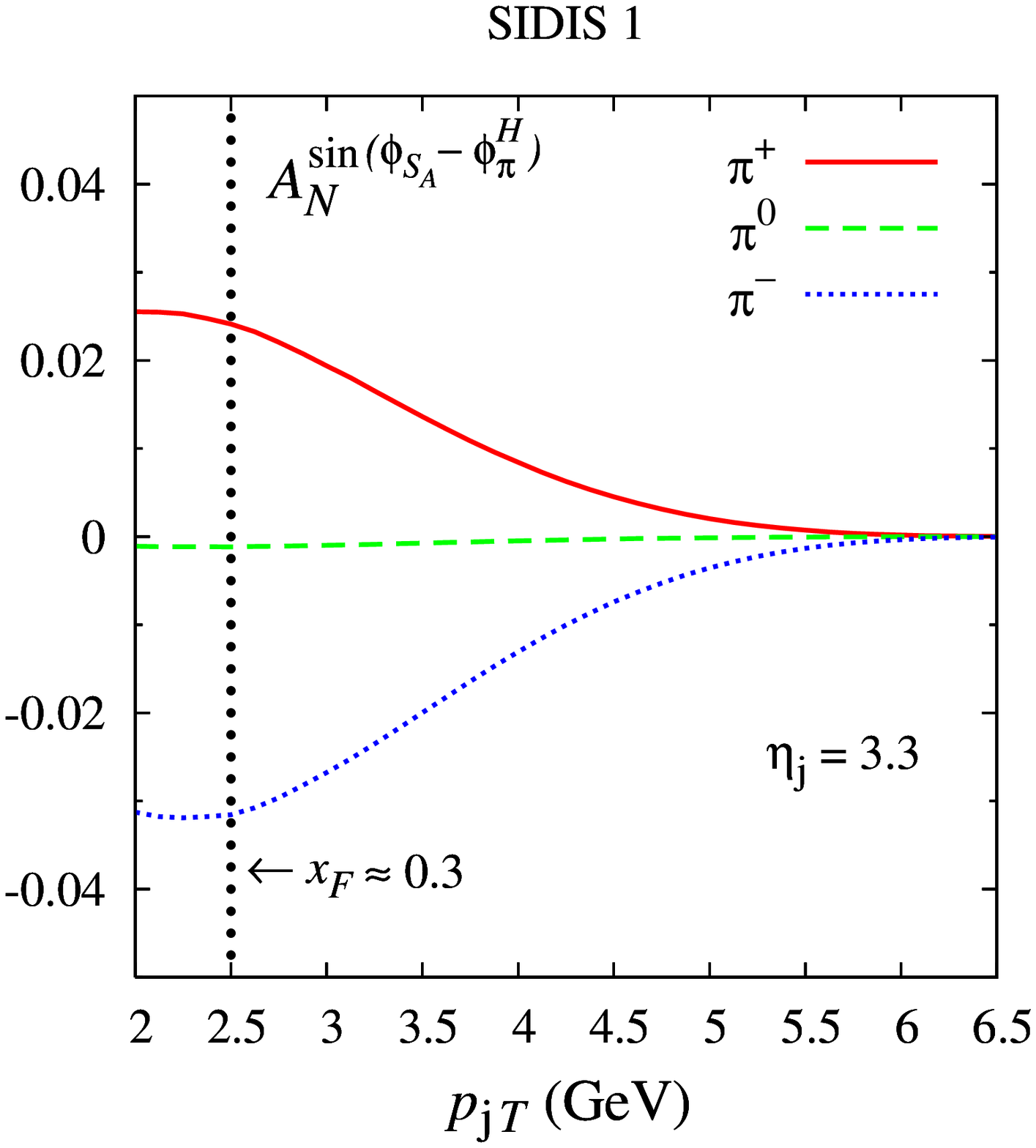}
 \includegraphics[angle=0,width=0.4\textwidth]{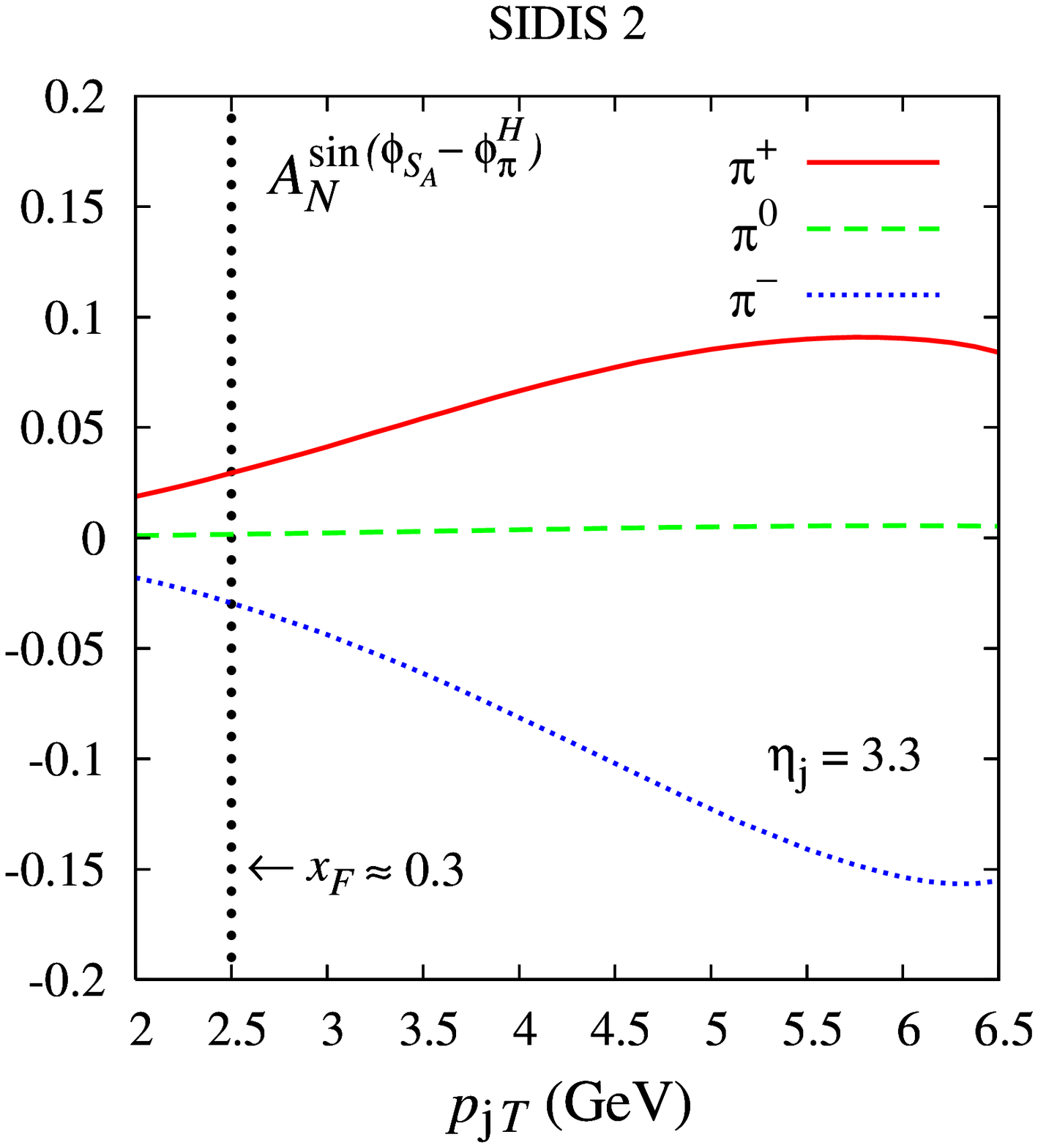}
 \caption{(color online). The estimated quark Collins asymmetry
for the $p^\uparrow p\to {\rm jet}+\pi + X$ process,
obtained adopting the parameterizations SIDIS~1 (left panel)
and SIDIS~2 (right panel) respectively,
at $\sqrt{s}=200$ GeV c.m.~energy in the forward rapidity
region and as a function of $p_{{\rm j}\,T}$,
from $p_{{\rm j}\,T}=2$ GeV up to the maximum allowed value.
Notice the difference in the scale
between the two panels.
The dotted black vertical line delimits the region
beyond which the SIDIS parameterizations for the quark
transversity distribution are presently plagued by large uncertainties.
Similar results are obtained when considering
different c.m.~energies.
\label{asy-an-coll-par200} }
\end{figure*}

Concerning the forward rapidity region,
in Fig.~\ref{asy-an-coll-par200} we present,
for both charged and neutral pions, some results
at $\sqrt{s}=200$ GeV c.m.~energy as a function of $p_{{\rm j}\,T}$,
adopting the SIDIS~1 (left panel) and the SIDIS~2
(right panel) parameterizations.
The Collins asymmetry for neutral pions (dashed green lines)
results to be practically negligible.
This can be easily understood since in the available parameterizations
the favoured (e.g.~$u\to\pi^+$) and unfavoured (e.g.~$d\to\pi^+$) Collins
fragmentation functions are comparable in size and opposite in sign.
Due to isospin symmetry the $\pi^0$ FFs are half the sum of those
for charged pions, therefore the $\pi^0$ Collins FF is always very small.
Moreover, the $u$, $d$ quark transversity distributions are also
opposite in sign, leading to additional cancellations among quark
contributions.
For charged pions, similarly to the case of the Sivers asymmetry, the two
parameterizations give comparable results (notice the
different scale adopted in the two panels)
in the $p_{{\rm j}\,T}$
region where the transversity distribution is reasonably
constrained by SIDIS data (see the dotted black vertical line),
while they lead to completely different estimates in the large
$p_{{\rm j}\,T}$ region where the parameterizations are
basically unconstrained by SIDIS data.
In particular, in this region the SIDIS~1 set gives almost
negligible results, while the SIDIS~2 set leads to
an asymmetry of about 8\%~($\pi^+$) and 15\%~($\pi^-$) in size,
which should be hopefully measurable. A measurement of this asymmetry
would be then very important and helpful in clarifying
the large $x$ behaviour of the quark transversity distribution.
A qualitatively similar situation is obtained at $\sqrt{s}=500$~GeV.

\subsubsection{\label{sec-results-an-jet}
Transverse single spin asymmetry for inclusive jet production}
For completeness we have extended our analysis to the transverse
single spin asymmetry for inclusive jet production in polarized
$pp$ collisions, $A_N(p^\uparrow p\to {\rm jet}+X)$.
In principle, this case can be obtained by the jet + pion
production process by integrating over the full pion phase space.
Of course, in this case the unobserved fragmentation process in the final
state plays no role in the azimuthal asymmetries, which
can only be originated by mechanisms, like the Sivers effect,
acting in the initial state.
Moreover, we have verified that for the kinematical configurations
considered in this paper all contributions but the Sivers effect
play a negligible role already in the maximized scenario.
Therefore, in what follows, we limit our discussion to the Sivers asymmetry.
As already mentioned, in this case
quark and gluon contributions cannot be disentangled since they
add up leading to a $\sin\phi_{S_A}$ asymmetry.

Let us first discuss the maximized scenario. In the central rapidity
region, the maximized gluon contribution is of the order 20\%
at the lowest $p_{{\rm j}\,T}$ values, decreasing fast to about 3\%
at large $p_{{\rm j}\,T}$ for all c.m.~energies considered.
The maximized quark contribution is of the order 1-3\% in the full
$p_{{\rm j}\,T}$ range, slowly decreasing with the increase of the
c.m.~energy.
The total potential effect is therefore sizable only at small
$p_{{\rm j}\,T}$ values due to the gluon component.
The situation is different in the forward rapidity region.
Here both quark and gluon maximized contributions can be
very sizable, showing as expected
an opposite, respectively increasing and decreasing,
 behaviour vs.~$p_{{\rm j}\,T}$. The total maximized
 Sivers effect is therefore large in the full $p_{{\rm j}\,T}$
 range with little dependence on the c.m.~energy.

Concerning numerical estimates obtained adopting the available
parameterizations SIDIS~1 and SIDIS~2 for the quark Sivers
function, and the updated bound on the gluon Sivers function,
the situation is the following:\\
1) In the central rapidity region, for both SIDIS~1,2 sets and
all energies considered the quark contribution is practically
negligible. Instead, the gluon contribution can be at most of the order
$10\div15$\% at the lowest $p_{{\rm j}\,T}$ values but decreases quickly
with the increasing of $p_{{\rm j}\,T}$. However, at least for
$\sqrt{s}=200$ and 500 GeV, it can still be about 2-4\% in the
upper $p_{{\rm j}\,T}$ range. The measurement of a comparable
Sivers asymmetry in these kinematical configurations could then
be a clear indication for a gluonic contribution to the Sivers effect.\\
2) In the forward rapidity region the quark contribution is small and
negative at $p_{{\rm j}\,T}=2$ GeV for both sets adopted, while at large
$p_{{\rm j}\,T}$ values it is negligible for the SIDIS~2 set
and positive and of the order 2-4\% for the SIDIS~1 set.
The gluon contribution can be sizable at very low $p_{{\rm j}\,T}$ values
but becomes negligible quickly as $p_{{\rm j}\,T}$ increases.

\begin{figure*}[t]
 \includegraphics[angle=0,width=0.4\textwidth]{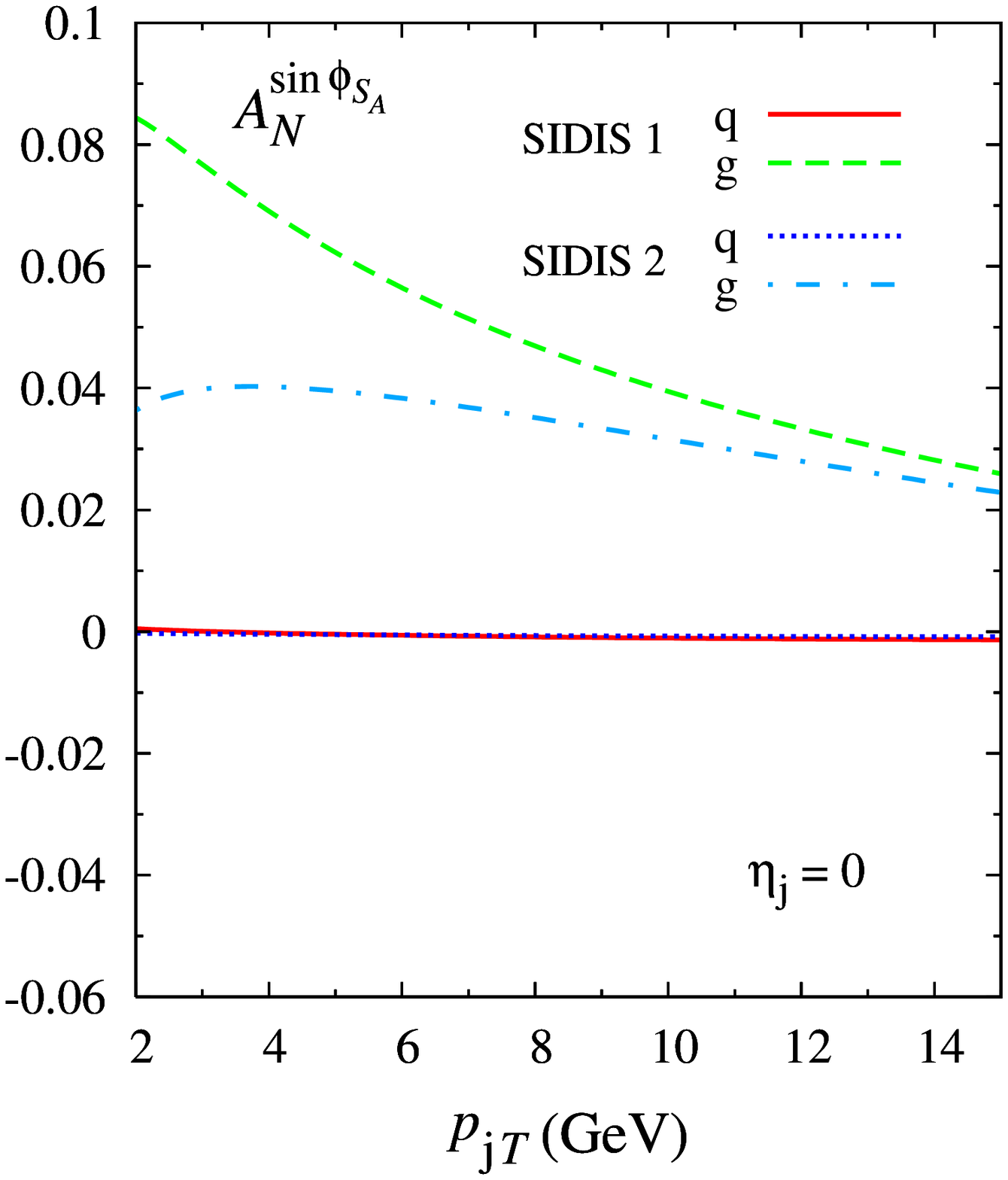}
 \includegraphics[angle=0,width=0.4\textwidth]{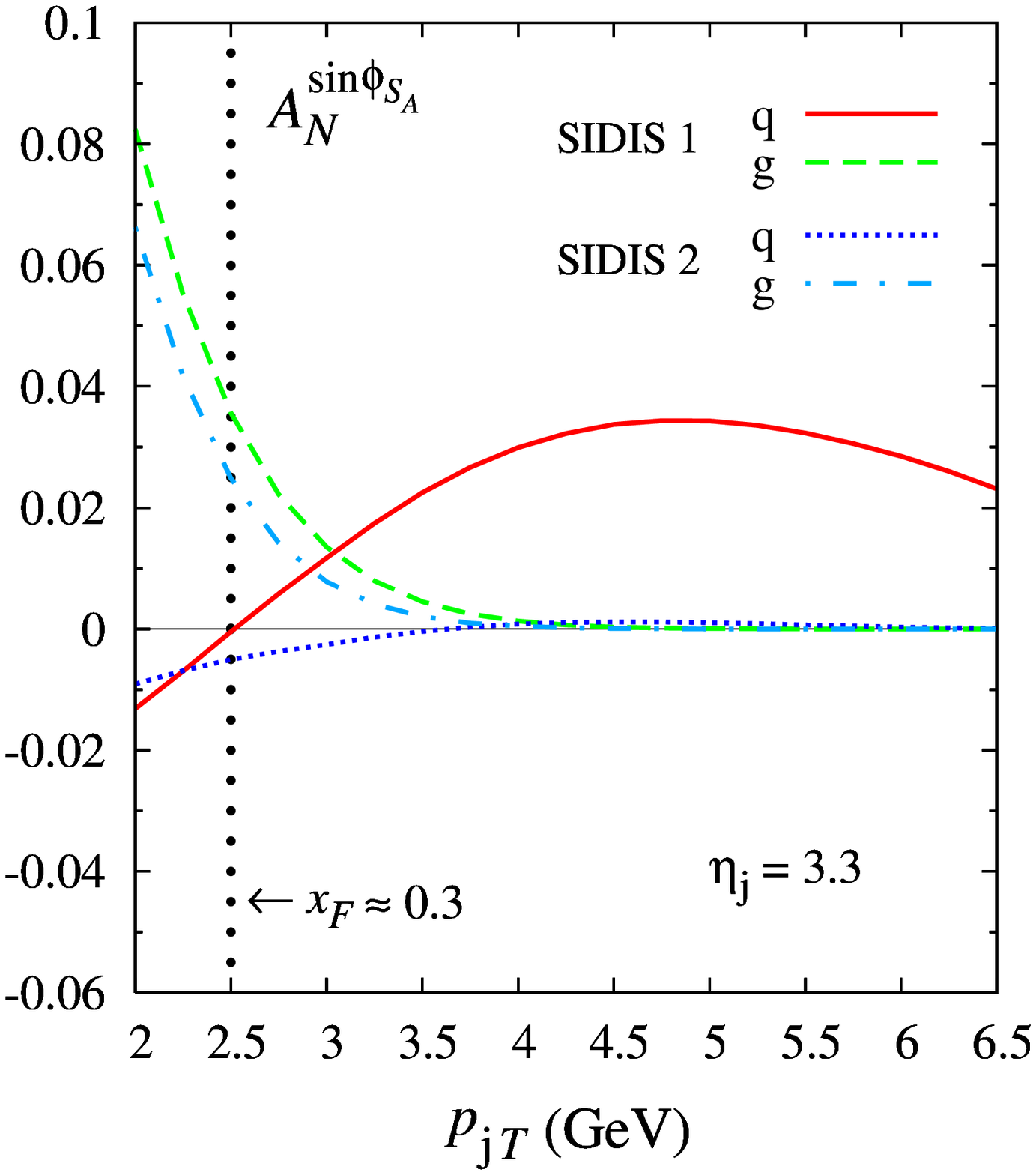}
 \caption{(color online).
 The estimated quark and gluon Sivers contributions to the
 transverse single spin asymmetry for the
 $p^\uparrow p\to{\rm jet} + X$ process,
at $\sqrt{s}=200$ GeV c.m.~energy in the central (left panel)
and forward (right panel) rapidity region as a function of $p_{{\rm j}\,T}$,
from $p_{{\rm j}\,T}=2$ GeV up to the maximum allowed value,
obtained adopting the parametrization sets
SIDIS~1 (quark contribution: solid red line;
gluon contribution: dashed green line) and SIDIS~2
(quark contribution: dotted blue line; gluon contribution:
dot-dashed cyan line).
The dotted black vertical line in the right panel delimits the region
beyond which the SIDIS parameterizations for the quark
transversity distribution are presently plagued by large uncertainties.
Similar results, with some differences
in the total size and in the relative weight of the quark and
gluon contributions are obtained
considering different c.m.~energies.
 \label{asy-siv-jet-200} }
\end{figure*}

As an example, in Fig.~\ref{asy-siv-jet-200} we show the estimated quark and
gluon Sivers contributions to the transverse single spin asymmetry
for inclusive jet production in the central (left panel) and
forward (right panel) rapidity regions at $\sqrt{s}=200$ GeV, obtained
adopting the SIDIS~1 and SIDIS~2 parameterizations for the quark
Sivers function and the updated bound for the gluon Sivers function
(assumed to be positive).

\section{\label{sec-conclusions} Conclusions}
In this paper we have presented a study of the azimuthal asymmetries measurable
in the distribution of leading unpolarized or spinless
hadrons (mainly pions) inside a large-$p_T$
jet produced in unpolarized and single-transverse polarized proton proton
collisions for kinematical configurations accessible at RHIC.
To this end, we have adopted a generalized TMD parton model approach
with inclusion of spin and intrinsic parton motion effects both
in the distribution and in the fragmentation sectors.
We have shown how a detailed phenomenological analysis of these
effects can be very useful in shedding light on several
aspects of azimuthal and transverse single spin asymmetries
in (un)polarized hadronic collisions.
It may also help in clarifying
the role played by the quark(gluon) Sivers distribution and by the Collins(-like)
fragmentation function in the sizable single spin asymmetries observed
at RHIC for forward pion production.
Available parameterizations for the TMD quark transversity and Sivers distribution
functions, obtained by fitting SIDIS and $e^+e^-$ data,
are presently largely unconstrained for
light-cone momentum fractions $x \geq 0.3$, that is the region
playing a fundamental role for forward pion production at RHIC.
The transverse single spin asymmetry for inclusive particle production
is a complicate higher-twist effect involving several TMD mechanisms
that cannot be easily disentangled as in the case of SIDIS and DY processes.
On the contrary, the leading-twist azimuthal asymmetries discussed in this
paper allow one, in close analogy with the SIDIS case, to discriminate
among different effects by taking suitable moments of the asymmetries.
Moreover, we have shown that in principle quark and gluon originating jets
can be distinguished, at least in some kinematical regimes.
Neglecting intrinsic motion in the distribution sector leaves at work only the Collins
azimuthal asymmetry.
As already shown by Yuan~\cite{Yuan:2007nd},
the measurements proposed in this paper would allow one to
determine unambiguously the role played by the Collins effect.
Universality properties of the Collins function have been proved, so that
this process can be complementary to the SIDIS and $e^+e^-$ measurements
in order to constrain the quark Collins fragmentation
function and, as an important by-product,
the large $x$ behaviour of the TMD transversity distribution function.

{}For the quark and gluon Sivers function and for the Boer-Mulders function
the situation is more complicate. Since factorization has not been proven yet
for inclusive particle production in hadronic collisions,
the use, based on universality, of the parameterizations
obtained by fitting SIDIS and $e^+e^-$data might be questionable.
The phenomenological analysis
proposed in this paper gives us the opportunity of testing the
factorization and universality assumptions, and of gaining
information on the size and \textit{sign} of the TMD functions discussed.
This can be very useful also for further developments of the TMD approach,
since it is difficult a priori to assess at which values of
the factorization scale the role and the size of possible
factorization-breaking terms become relevant and non negligible.

Let us finally stress again that the unambiguous measurement of any of the
asymmetries, other than the Collins one, discussed in this paper would be a clear indication
of the role played by intrinsic parton motion in the initial colliding
hadrons for the spin asymmetry sector in polarized hadronic collisions.
\begin{acknowledgments}
We are grateful to Pietro Contu and Mattia Melis for collaboration during the early stages
of this work.
C.P.~is supported by Regione Autonoma della Sardegna (RAS) through a research
grant under the PO Sardegna FSE 2007-2013, L.R. 7/2007, ``Promozione della
ricerca scientifica e dell'innovazione tecnologica in Sardegna".
U.D.~and F.M.~acknowledge partial support by Italian Ministero dell'Istruzione,
dell'Universit\`{a} e della Ricerca Scientifica (MIUR) under Cofinanziamento PRIN 2008,
and by the European Community under the FP7 ``Integrating Activities" project
``Study of Strongly Interacting Matter" (HadronPhysics2), grant agreement No.~227431.
\end{acknowledgments}
%

%
\end{document}